\documentclass[aps,prd,nofootinbib,twocolumn,superscriptaddress,preprintnumbers,balancelastpage,longbibliography,nobibnotes,preprintnumbers]{revtex4-1}
\usepackage{graphicx,epsfig,psfrag,bm,amssymb,adjustbox}
\usepackage{dcolumn}
\usepackage{bm}
\usepackage[dvipsnames]{xcolor}
\usepackage{braket}
\usepackage{mathrsfs,amsfonts,hepunits,color}
\usepackage{hyperref}
\usepackage{comment}
\usepackage{physics}
\usepackage{subcaption}
\usepackage[labelfont=bf,
   justification=justified,
   format=plain]{caption}

\newcommand{\E}{\mathbf{E}}
\newcommand{\B}{\mathbf{B}}
\newcommand{\Pol}{\mathbf{P}}
\newcommand{\M}{\mathbf{M}}
\newcommand{\J}{\mathbf{J}}
\newcommand{\g}{g_{a \gamma \gamma}}

\newcommand{\Ap}{A^{\prime}}

\newcommand{\rhodm}{\rho_{_\text{DM}}}

\begin{document}

\title{Searches for New Particles, Dark Matter,\\ and Gravitational Waves with SRF Cavities}
\affiliation{Superconducting Quantum Materials and Systems Center (SQMS), Fermi National Accelerator Laboratory, Batavia, IL 60510, USA}
\author{Asher~Berlin}
\affiliation{Theory Division, Fermi National Accelerator Laboratory, Batavia, IL 60510, USA}
\affiliation{Superconducting Quantum Materials and Systems Center (SQMS), Fermi National Accelerator Laboratory, Batavia, IL 60510, USA}
%
\author{Sergey~Belomestnykh}
\affiliation{Superconducting Quantum Materials and Systems Center (SQMS), Fermi National Accelerator Laboratory, Batavia, IL 60510, USA}
\affiliation{Applied Physics and Superconducting Technology Division, Fermi National Accelerator Laboratory, Batavia, IL 60510, USA}
\affiliation{Department of Physics and Astronomy, Stony Brook University, Stony Brook, NY 11794, USA}
\author{Diego~Blas}
\affiliation{Grup de F\'isica Te\`orica, 
Departament  de  F\'isica, Universitat  Aut\`onoma  de  Barcelona,   Bellaterra, 08193 Barcelona, Spain}
\affiliation{Institut de Fisica d’Altes Energies (IFAE), The Barcelona Institute of Science and Technology,\\ Campus UAB, 08193 Bellaterra  (Barcelona), Spain}
\author{Daniil~Frolov}
\affiliation{Superconducting Quantum Materials and Systems Center (SQMS), Fermi National Accelerator Laboratory, Batavia, IL 60510, USA}
\author{Anthony~J.~Brady}
\affiliation{
Department of Electrical and Computer Engineering, University of Arizona, Tucson, Arizona 85721, USA}
\affiliation{Superconducting Quantum Materials and Systems Center (SQMS), Fermi National Accelerator Laboratory, Batavia, IL 60510, USA}
\author{Caterina~Braggio}
\affiliation{Dipartimento di Fisica e Astronomia, Padova, Italy}
\affiliation{INFN, Sezione di Padova, Padova, Italy}
\affiliation{Superconducting Quantum Materials and Systems Center (SQMS), Fermi National Accelerator Laboratory, Batavia, IL 60510, USA}
\author{Marcela~Carena}
\affiliation{Theory Division, Fermi National Accelerator Laboratory, Batavia, IL 60510, USA}
\affiliation{Department of Physics, University of Chicago, Chicago, Illinois, 60637, USA}
\affiliation{Superconducting Quantum Materials and Systems Center (SQMS), Fermi National Accelerator Laboratory, Batavia, IL 60510, USA}
\author{Raphael~Cervantes}
\affiliation{Superconducting Quantum Materials and Systems Center (SQMS), Fermi National Accelerator Laboratory, Batavia, IL 60510, USA}
\author{Mattia~Checchin}
\affiliation{Superconducting Quantum Materials and Systems Center (SQMS), Fermi National Accelerator Laboratory, Batavia, IL 60510, USA}
\author{Crispin~Contreras-Martinez}
\affiliation{Superconducting Quantum Materials and Systems Center (SQMS), Fermi National Accelerator Laboratory, Batavia, IL 60510, USA}
\affiliation{Applied Physics and Superconducting Technology Division, Fermi National Accelerator Laboratory, Batavia, IL 60510, USA}
\author{Raffaele~Tito~D'Agnolo}
\affiliation{Universit\'e Paris Saclay, CEA, CNRS, Institut de Physique Th\'eorique, 91191, Gif-sur-Yvette, France}
\author{Sebastian~A.~R.~Ellis}
\affiliation{D\'epartement de Physique Th\'eorique, Universit\'e de Gen\`eve,\\
24 quai Ernest Ansermet, 1211 Gen\`eve 4, Switzerland}
\author{Grigory~Eremeev}
\affiliation{Superconducting Quantum Materials and Systems Center (SQMS), Fermi National Accelerator Laboratory, Batavia, IL 60510, USA}
\affiliation{Applied Physics and Superconducting Technology Division, Fermi National Accelerator Laboratory, Batavia, IL 60510, USA}
\author{Christina~Gao}
\affiliation{Department of Physics, University of Illinois at Urbana-Champaign, Urbana, IL 61801, USA}
\affiliation{Theory Division, Fermi National Accelerator Laboratory, Batavia, IL 60510, USA}
\affiliation{Superconducting Quantum Materials and Systems Center (SQMS), Fermi National Accelerator Laboratory, Batavia, IL 60510, USA}
\author{Bianca~Giaccone}
\affiliation{Superconducting Quantum Materials and Systems Center (SQMS), Fermi National Accelerator Laboratory, Batavia, IL 60510, USA}
\author{Anna~Grassellino}
\affiliation{Superconducting Quantum Materials and Systems Center (SQMS), Fermi National Accelerator Laboratory, Batavia, IL 60510, USA}
\affiliation{Applied Physics and Superconducting Technology Division, Fermi National Accelerator Laboratory, Batavia, IL 60510, USA}
\author{Roni~Harnik}
\email{roni@fnal.gov}
\affiliation{Superconducting Quantum Materials and Systems Center (SQMS), Fermi National Accelerator Laboratory, Batavia, IL 60510, USA}
\affiliation{Theory Division, Fermi National Accelerator Laboratory, Batavia, IL 60510, USA}
\author{Matthew~Hollister}
\affiliation{Superconducting Quantum Materials and Systems Center (SQMS), Fermi National Accelerator Laboratory, Batavia, IL 60510, USA}
\affiliation{Applied Physics and Superconducting Technology Division, Fermi National Accelerator Laboratory, Batavia, IL 60510, USA}
\author{Ryan~Janish}
\affiliation{Theory Division, Fermi National Accelerator Laboratory, Batavia, IL 60510, USA}
\affiliation{Superconducting Quantum Materials and Systems Center (SQMS), Fermi National Accelerator Laboratory, Batavia, IL 60510, USA}
\author{Yonatan~Kahn}
\affiliation{Department of Physics, University of Illinois at Urbana-Champaign, Urbana, IL 61801, USA}
\affiliation{Superconducting Quantum Materials and Systems Center (SQMS), Fermi National Accelerator Laboratory, Batavia, IL 60510, USA}
\author{Sergey~Kazakov}
\affiliation{Superconducting Quantum Materials and Systems Center (SQMS), Fermi National Accelerator Laboratory, Batavia, IL 60510, USA}
\affiliation{Applied Physics and Superconducting Technology Division, Fermi National Accelerator Laboratory, Batavia, IL 60510, USA}
\author{Doga~Murat~Kurkcuoglu}
\affiliation{Fermilab Quantum Institute, Fermi National Accelerator Laboratory, Batavia, IL 60510, USA}
\affiliation{Superconducting Quantum Materials and Systems Center (SQMS), Fermi National Accelerator Laboratory, Batavia, IL 60510, USA}
\author{Zhen~Liu}
\affiliation{School of Physics and Astronomy, University of Minnesota, Minneapolis, MN 55455, USA}
\affiliation{Superconducting Quantum Materials and Systems Center (SQMS), Fermi National Accelerator Laboratory, Batavia, IL 60510, USA}
\author{Andrei~Lunin}
\affiliation{Superconducting Quantum Materials and Systems Center (SQMS), Fermi National Accelerator Laboratory, Batavia, IL 60510, USA}
\author{Alexander~Netepenko}
\affiliation{Superconducting Quantum Materials and Systems Center (SQMS), Fermi National Accelerator Laboratory, Batavia, IL 60510, USA}
\affiliation{Applied Physics and Superconducting Technology Division, Fermi National Accelerator Laboratory, Batavia, IL 60510, USA}
\author{Oleksandr~Melnychuk}
\affiliation{Superconducting Quantum Materials and Systems Center (SQMS), Fermi National Accelerator Laboratory, Batavia, IL 60510, USA}
\author{Roman Pilipenko}
\affiliation{Superconducting Quantum Materials and Systems Center (SQMS), Fermi National Accelerator Laboratory, Batavia, IL 60510, USA}
\affiliation{Applied Physics and Superconducting Technology Division, Fermi National Accelerator Laboratory, Batavia, IL 60510, USA}
\author{Yuriy~Pischalnikov}
\affiliation{Superconducting Quantum Materials and Systems Center (SQMS), Fermi National Accelerator Laboratory, Batavia, IL 60510, USA}
\affiliation{Applied Physics and Superconducting Technology Division, Fermi National Accelerator Laboratory, Batavia, IL 60510, USA}
\author{Sam Posen}\email{sposen@fnal.gov}
\affiliation{Superconducting Quantum Materials and Systems Center (SQMS), Fermi National Accelerator Laboratory, Batavia, IL 60510, USA}
\affiliation{Applied Physics and Superconducting Technology Division, Fermi National Accelerator Laboratory, Batavia, IL 60510, USA}
\author{Alex~Romanenko}
\affiliation{Superconducting Quantum Materials and Systems Center (SQMS), Fermi National Accelerator Laboratory, Batavia, IL 60510, USA}
\affiliation{Applied Physics and Superconducting Technology Division, Fermi National Accelerator Laboratory, Batavia, IL 60510, USA}
\author{Jan~Schütte-Engel}
\affiliation{Department of Physics, University of Illinois at Urbana-Champaign, Urbana, IL 61801, USA}
\affiliation{Superconducting Quantum Materials and Systems Center (SQMS), Fermi National Accelerator Laboratory, Batavia, IL 60510, USA}
\author{Changqing~Wang}
\affiliation{Superconducting Quantum Materials and Systems Center (SQMS), Fermi National Accelerator Laboratory, Batavia, IL 60510, USA}
\author{Vyacheslav~Yakovlev}
\affiliation{Superconducting Quantum Materials and Systems Center (SQMS), Fermi National Accelerator Laboratory, Batavia, IL 60510, USA}
\affiliation{Applied Physics and Superconducting Technology Division, Fermi National Accelerator Laboratory, Batavia, IL 60510, USA}
\author{Kevin Zhou}
\affiliation{SLAC National Accelerator Laboratory, 2575 Sand Hill Road, Menlo Park, CA 94025, USA}
\author{Silvia~Zorzetti}
\affiliation{Superconducting Quantum Materials and Systems Center (SQMS), Fermi National Accelerator Laboratory, Batavia, IL 60510, USA}
\author{Quntao~Zhuang} 
\affiliation{
Department of Electrical and Computer Engineering, University of Arizona, Tucson, Arizona 85721, USA}
\affiliation{Superconducting Quantum Materials and Systems Center (SQMS), Fermi National Accelerator Laboratory, Batavia, IL 60510, USA}
\affiliation{
J. C. Wyant College of Optical Sciences, University of Arizona, Tucson, Arizona 85721, USA}


\date{\today}

\begin{abstract}
This is a Snowmass white paper on the utility of existing and future superconducting  cavities to probe fundamental physics. Superconducting radio frequency (SRF) cavity  technology has seen tremendous progress in the past decades, as a tool for accelerator science. With advances spearheaded by the SQMS center at Fermilab, they are now being brought to the quantum regime becoming a tool in quantum science thanks to the high degree of coherence. The same high quality factor can be leveraged in the search for new physics, including searches for new particles, dark matter, including the QCD axion, and gravitational waves. We survey some of the physics opportunities and the required directions of R\&D. Given the already demonstrated integration of SRF cavities in large accelerator systems, this R\&D may enable larger scale searches by dedicated experiments.
\begin{center}
---\\
\emph{Submitted to the  Proceedings of the US Community Study}\\ 
\emph{on the Future of Particle Physics (Snowmass 2021)}\\ ---
\end{center}
\end{abstract}

\preprint{ \hspace{12cm}FERMILAB-PUB-22-150-SQMS-T\  \ $\vcenter{\includegraphics[width=01.5cm]{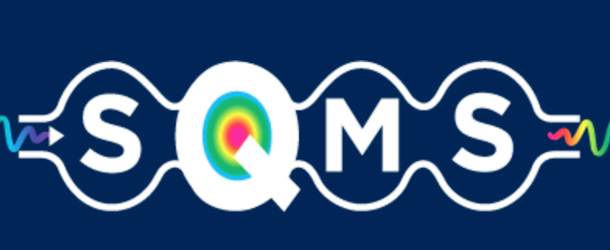}}$}

\maketitle

\tableofcontents

\section{Introduction}

Superconducting radio frequency (SRF) cavities are critical components in particle accelerators. When excited at their resonant frequency, they build up large-amplitude electric fields while dissipating very little heat. This makes it possible for them to provide extremely efficient conversion of radio-frequency power into energy gain for the charged particle beams that pass through them. But SRF cavities are, in their essence, extremely high quality electromagnetic resonators, devices that are now of strong active interest for quantum information science (QIS). For quantum computing, quantum states can be stored and manipulated in electromagnetic resonators, and superconductors at millikelvin temperatures are employed to sustain the coherence of the quantum states for long enough to perform complex computations. For quantum sensing, SRF cavities can furnish a large volume where very weak signals of radio-frequency photons can be collected, with only a small fraction of photons being lost to heat at the cavity walls.

Accelerator scientists have been improving SRF cavities for 50 years, and new accelerator applications have been enabled as performance has continuously improved over the years~\cite{CEBAF, SNS, XFEL, LCLSII}. The performance improvements were the results of improved understanding of RF superconductivity and materials science as well as new SRF cavity processing techniques that were developed to overcome performance-limiting phenomena, such as field emission, and to enhance superconducting properties~\cite{electropolishing, HPR, HPR2, 120Cbake, 120Cbake2, nitrogen-doping, 2-step-bake, midTbake}.  The advances in cavity performance, and the modern understanding of SRF systems that has been developed for particle accelerator applications, can also be leveraged for QIS. In this whitepaper, we focus on quantum sensing applications of SRF cavities.

\subsection{Dark Matter vs. New Particle Searches}

The search for new physics follows the standard scientific paradigm of hypothesis testing. In the context of light degrees of freedom that couple to photons, there are two distinct hypotheses that can be tested:
\paragraph{\bf A new particle:} 
It is well motivated to test the notion that there may be new degrees of freedom in the Lagrangian (or Hamiltonian) which interacts with Standard Model (SM) particles. The well-motivated cases that will be best probed by electromagnetic cavities are new particles that interact with photons, such as axions or dark photons. Hence, examples of hypotheses that we can test are
\begin{equation}
\mbox{Hypothesis:}\qquad    \mathcal{L} \supset \mbox{ axion? dark photon?}
\end{equation}
where $\mathcal{L}$ is the Lagrangian density.
Experiments that test this hypothesis produce and detect the new particle. As such, the experimenter has control over the frequency of the axion or dark photon field, as well as over its phase. Light-shining-through-wall (LSW) experiments~\cite{Okun:1982xi,VanBibber:1987rq,Hoogeveen:1990vq} are a canonical example of axion and dark photon searches in this class.\footnote{Collider and fixed-target experiments are also in this category, but these typically search for heavier and more strongly-coupled particles.} Searches for high-frequency gravitational waves (GWs), while not strictly speaking a test for new particles (at least since direct detection of GWs at LIGO~\cite{LIGOScientific:2021djp}), are closely related to new particle searches because gravity couples similarly to the photon as do axions. 

\paragraph{\bf A dark matter candidate:}
 Here we test the notion that a new set of degrees of freedom exist, as above, and in addition these degrees of freedom make up the dark matter (DM) that dominates our galaxy. Hence, the tested hypothesis is

\begin{equation} 
\mbox{Hypothesis:}\ \
\left\{ \begin{array}{c}
     \qquad\ \mathcal{L}\ \ \supset\ \mbox{ axion? dark photon?}\\[8pt]
   \qquad \mbox{and}\\[7pt]
   \vcenter{\includegraphics[width=.7cm]{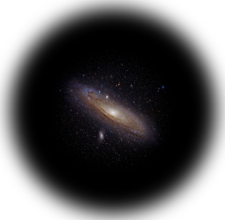}} \hspace{-7.4
   cm} \supset \mbox{ axion? dark photon?}
\end{array}\right.
\end{equation}
where the DM halo is represented pictorially. As opposed to pure lab searches, we do not know the frequency of the DM signal nor its phase. Because the DM is non-relativistic, it is expected to have a velocity of order the galactic virial velocity $v\sim 10^{-3} \, c$. The frequency $\omega$ of the DM field is thus set by the DM mass, an unknown parameter, with a small spread $\Delta \omega$, of order $\Delta \omega/\omega \sim v^2\sim 10^{-6}$. The wave-like DM field can therefore be thought of as a coherent classical source with an unknown frequency and an effective quality factor of order $Q_\mathrm{DM}\sim 10^6$. \\

We note that these two hypotheses are complementary and can be pursued somewhat independently. Of course, the DM candidate hypothesis is strictly stronger than the new particle hypothesis. 
A new particle search, such as an axion LSW experiment, is not strictly constrained by an axion DM search, because the axion need not be the DM.
On the other hand, a null result in the parameter space of a new particle search will necessarily constrain the parameter space of a DM search. That said, a DM search is likely to have an advantage in probing the feeble couplings of a new particle because the number density of DM particles in the regime of interest is typically much larger than the density of particles that can feasibly be produced in the lab. 

\section{Searches for New Degrees of Freedom}

Here we summarize opportunities to search for new particles with SRF cavities. The central characters are dark photons and axions (or axion-like particles), though we will touch upon other new physics candidates. We will discuss dark photons and axions in turn.

\subsection{Dark Photon Searches}

The dark photon~\cite{Holdom:1985ag} is a well-motivated extension of the SM, with its allure lying  in its simplicity and theoretical completeness. The model posits that in addition to the photon there is another $U(1)$ gauge field, the dark photon. The dark photon, unlike its SM counterpart, is massive (which breaks the new $U(1)$ symmetry). The only renormalizable interaction (and hence the only relevant interaction at low energies) one can write involving the dark photon is a kinetic mixing with the SM photon. The Lagrangian is 
\begin{eqnarray}
\mathcal{L} = \mathcal{L}_\mathrm{SM} + \epsilon \, F^{\mu\nu}F'_{\mu\nu} + \frac{1}{2} \, m_{A'}^2 \, A'_\mu A'^\mu
~,
\label{eq:darkphotonL}
\end{eqnarray}
where primes adorn the dark photon field strength and vector potential and $m_{A'}$ is the dark photon mass.

The dark photon is constrained by a host of experiments including collider and fixed-target setups at relatively high masses, as well as stars and CMB observations (see, e.g., Ref.~\cite{Essig:2013lka} for a review). As a rule, limits on dark photons become weaker as a power of the dark photon mass over the frequency it is emitted in a given experiment. This is because the kinetic mixing term in Eq.~(\ref{eq:darkphotonL}) can be rotated away by a field redefinition in the absence of a mass term. It has been shown that limits on the longitudinal polarization of dark photons are weakened more slowly at low mass and are hence more constraining~\cite{An:2013yfc, An:2013yua}. 

LSW experiments have set new limits on dark photons in optical setups, such as OSQAR~\cite{Ballou:2015cka} and ALPS~\cite{Hoogeveen:1990vq, Sikivie:2007qm, Ehret:2009sq} with enhanced plans for ALPS~II~\cite{Bahre:2013ywa}, and in RF setups such as CROWS~\cite{Betz:2013dza}. To date, these experiments have not taken advantage of the enhanced sensitivity to longitudinal dark photons.

\subsection{Light-shining-through wall: Dark SRF}


We can explore dark photon scenarios using SRF cavities in a setup that resembles shining light through a cavity wall. The conversion of some of the photons to dark photons before the wall and conversion back to regular photons past the wall makes such a detection possible, if dark photons exist at a hypothesized mass and coupling.  Resonant cavities can be used on both sides of the wall to increase the number of photons on the emitting side and to enhance the probability of conversion of dark photons to visible ones on the receiver side. In particular, in an RF cavity the system can be designed to search for the parametrically enhanced longitudinal coupling of the dark photon~\cite{Graham:2014sha}. The Dark SRF experiment plans to conduct such a search with ultra-high quality cavities~\cite{DarkSRF,DarkSRF2, DarkSRFpaper}.

\begin{figure}[t]
 \includegraphics[scale=0.7]{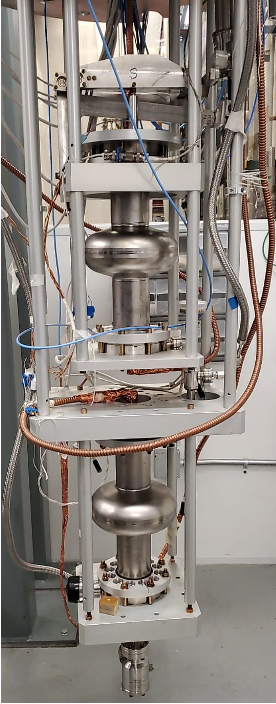}
\caption{\label{fig:DarkSRFpic}The experimental setup for the Dark SRF experiment consisting of two 1.3~GHz cavities~\cite{DarkSRFpaper}.
}
\end{figure}

\begin{figure*}[t]
 \includegraphics[width=0.65\linewidth]{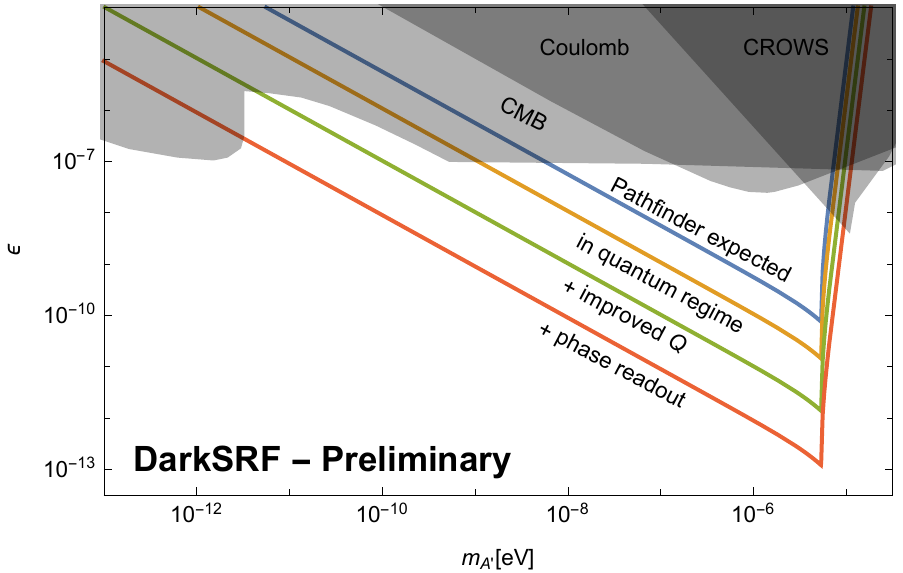}
\caption{\label{fig:Exclusion} The projected 95\% C.L. sensitivity on the dark photon parameter space~\cite{DarkSRFpaper} from an upcoming pathfinder experiment (blue) along with various future SQMS focus targets (orange, green, and red). 
Current limits from the cavity experiment CROWS~\cite{Betz:2013dza}, 
a Cavendish-type test of Coulomb's law~\cite{Williams:1971ms}, and CMB observations are shown as gray shaded regions~\cite{Mirizzi:2009iz, Mirizzi:2009nq, Caputo:2020bdy}.}
\end{figure*}

The kinetic mixing term in Eq.~(\ref{eq:darkphotonL}) effectively allows the SM and dark electromagnetic fields to mix. As a result, a coherently oscillating SM electromagnetic field acts as a source of dark photons at the same frequency. In addition, a coherent dark photon field can resonantly excite an RF cavity with the appropriate frequency. Given an emitter cavity with quality factor $Q$ pumped with a power $P_\mathrm{em}$ at a frequency $\omega$, the radiated dark photon field will then deposit power in a nearby receiver cavity
\begin{equation}
    P_\mathrm{rec}=\epsilon^4 \left(\frac{m_{A'}}{\omega}\right)^4 \left|G\right|^2 Q^2 P_\mathrm{em}
    ~,
    \label{eq:sig}
\end{equation}
where $G$ is a form factor that is specified in the Appendix, but heuristically consists of the wavefunction overlap of the dark photon field with the spatial mode shape in the receiver cavity.
Assuming that the noise in the receiver is thermal, the signal-to-noise ratio (SNR) is given by the radiometer formula~\cite{Dicke:1946glx}
\begin{equation}
    \mathrm{SNR}=\frac{P_\mathrm{rec}}{P_\mathrm{th}}
    \sqrt{\delta\nu\, t_\mathrm{int}} =
    \frac{P_\mathrm{rec}}{k_B T_\mathrm{eff}}
    \sqrt{\frac{t_\mathrm{int}}{\delta\nu}}
    ~,
    \label{eq:SNR}
\end{equation}
where $\delta\nu$ is the bandwidth of the analysis, $t_\mathrm{int}$ is the integration time, and $T_\mathrm{eff}$ is the effective noise temperature. The factor of $\sqrt{\delta\nu\, t_\mathrm{int}}$ is the root number of independent measurements.

The experimental setup shown in Fig.~\ref{fig:DarkSRFpic} has been assembled using two 1.3~GHz high quality factor $Q_0$ SRF cavities and is employed in the current run of the Dark SRF experiment at SQMS. The top cavity is the emitter cavity and the bottom cavity is the receiver cavity. This setup is a powerful dark photon experiment provided that it properly matches frequencies, minimizes amplifier noise, and reduces cross talk between cavities. 
Detailed technical information will be presented in an upcoming work of Dark SRF~\cite{DarkSRFpaper}.

In Fig.~\ref{fig:Exclusion} we show the expected reach of an upcoming Dark SRF pathfinder run as well as future projections meeting various SQMS targets. The pathfinder run will be sensitive to regions above the blue solid line. Compared to existing constraints from the cavity experiment CROWS~\cite{Betz:2013dza}, a test of Coulomb's law~\cite{Williams:1971ms}, and observations of the CMB~\cite{Mirizzi:2009iz, Mirizzi:2009nq, Caputo:2020bdy}, this search will provide the world's best limit on dark photons in this mass range, improving constraints on $\epsilon$ by up to two orders of magnitude for dark photon masses between $10^{-8}$ and $6\times10^{-6}$~eV. We note that the improvement in $\epsilon$ sensitivity scales as ${\rm SNR}^{\mathrm 1/4}$, which clearly shows that the Dark SRF setup is advantageous in many aspects. Main factors contributing to such improvement are: high $Q$, noise suppression, and optimal placement of the receiver cavity. The pathfinder setup corresponds to approximately $\order{1000}$ photons. Enhancing this sensitivity to the few-photon limit in the receiver cavity requires lower temperatures (achievable with a dilution fridge) and improved suppression in amplifier noise. One can also improve the quality factors of both the emitter and receiver cavities to probe deeper into the dark photon parameter space. Finally, a phase sensitive readout of the receiver cavity will enable further improvements. These active research areas will be developed in the coming years and their corresponding projections are shown as additional colored lines in Fig.~\ref{fig:Exclusion}. 


The longitudinal Dark SRF setup is also sensitive to other new physics. In particular, the high $E$-fields in the emitter cavity can cause Schwinger pair-production of very light millicharged particles (mCPs)~\cite{Berlin:2020pey}. Upon production in the emitter cavity, these MCPs will then be accelerated in the oscillating RF fields, and due to the longitudinal  arrangement of the cavities will pass through the receiver cavity. This process  occurs with a period that is set by the emitter cavity frequency and hence is in resonance with the receiver. Ref.~\cite{Berlin:2020pey} has shown that this setup can set the strongest lab limits on mCPs for masses below $10^{-4}$~eV.

\subsection{Proposals for Axion Searches}

Axions and axion-like particles (ALPs) are  natural and well-motivated extensions of the SM.  The QCD axion was proposed as a natural solution of the strong CP problem~~\cite{PhysRevLett.38.1440,PhysRevD.16.1791,PhysRevLett.40.223,PhysRevLett.40.279}. It is a pseudo Nambu-Goldstone boson, arising from the spontaneous symmetry breaking (SSB) of an approximate global $U(1)$, and thus is naturally light. The solution to strong CP is possible due to the axion's coupling to gluons. Once this coupling is posited, a coupling to photons is a natural prediction of the model,
\begin{equation}
\label{eq:L_axion}
    \mathcal{L} \supset \frac{1}{4} \, g_{a\gamma \gamma}\, a\, F\tilde F = g_{a\gamma \gamma}\, a\, \vec E \cdot \vec B \,. 
\end{equation}
The ALP is  a generalization of the QCD axion which does not couple to QCD, but does couple to photons or SM fermions. 
ALPs are well motivated in their own right in top-down constructions~\cite{Svrcek:2006yi, Arvanitaki:2009fg}. 

The coupling in Eq.~(\ref{eq:L_axion}) is nonlinear. As a result, such an axion can mix with a photon in the presence of a background electromagnetic field. LSW experiments are a classic laboratory setup to search for ALPs.\footnote{ALP searches that are not LSW experiments include PVLAS~\cite{DellaValle:2015xxa}, a search for vacuum birefringence in a magnetic field. A different recently proposed example is the ``dark SPDC'' process, in which an optical pump photon down-converts to a signal and an axion (missing energy)~\cite{Estrada:2020dpg}.} In these experiments, a large number of photons is kept in an enclosed region with a strong constant background magnetic field, corresponding to the emitter cavity. In the presence of the $g_{a\gamma \gamma}$ interaction,  photons will convert to axions and escape the enclosure. A receiver cavity with a similarly strong field is set up nearby to detect axions that convert back to photons.
The current best limit from such LSW setups has been achieved by the OSQAR experiment: $|g_{a\gamma \gamma}|<3.5\times 10^{-8}\ \textrm{GeV}^{-1} $ for $m_a<0.3$~eV~\cite{Ballou:2015cka}. 
Operating at optical frequencies with high-finesse cavities, the ALPS experiment took advantage of resonant production and detection~\cite{Hoogeveen:1990vq, Sikivie:2007qm} and achieved a limit~\cite{Ehret:2009sq} comparable to that set by OSQAR, with the prospect of an improvement by a factor $\sim10^3$ in ALPS II~\cite{Bahre:2013ywa}. 
At microwave frequencies, the LSW setup has been implemented by the CROWS experiment and has set a comparable limit~\cite{Betz:2013dza}.

Like the dark photon case, LSW-type axion searches can benefit from high quality factors, which warrants the harnessing of advances in  SRF technology. The necessity for a background magnetic field, however presents a challenge, as high-quality superconductivity does not survive large fields. To overcome this, several proposals have been put forth, as we discuss below.  

\subsubsection{Two cavities with Static Field}

  \begin{figure} [t]
   \includegraphics[width=\columnwidth]{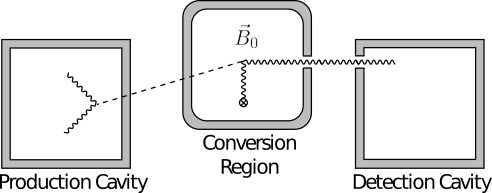}
   \caption 
   { \label{fig:LSWschematic} 
   Schematic depiction of an SRF light-shining-though-wall cavity search for axions~\cite{Janish:2019dpr}. The conversion region is in practice a superconducting toroidal magnet and the coupling to the detector (i.e., receiver) cavity is achieved via a pickup loop in the center of the toroid.  
   }
   \end{figure} 
   
One technique to utilize both high-$Q$ SRF cavities and large magnetic fields for a LSW axion search is to sequester the required magnetic fields away from the production and detection cavities.  
This was studied in Ref.~\cite{Janish:2019dpr} and is depicted schematically in Fig.~\ref{fig:LSWschematic}.
The experiment consists of three primary components: two SRF cavities and a conversion region containing a confined magnetic field.
Source modes are driven so that $\vec{E}\cdot\vec{B}$ is non-vanishing in the emitter cavity, which allows for the possibility of axion production. This is in contrast to the conventional RF approach of applying a static magnetic field to the emitter cavity, which for an SRF cavity would result in magnetic flux vorticies in the cavity walls that readily dissipate RF currents and spoil the cavity's large $Q$-factor.
Produced axions will pass through the cavity walls and propagate to the conversion region. 
The conversion region is a volume of confined magnetic field in the form of a superconducing, broken-toroid magnet. 
In this magnetic field some axions will convert back to photons. 
It is essential that these photons are not confined in the conversion region as is the static magnetic field, but rather they must propagate to a receiver which couples them into the receiver cavity.  
This is achieved by opening gaps in the toroidal structure of the magnet, for example by constructing it from polodially wrapped superconducting wire.
This forms an effective polarization filter which confines the toroidal-pointing static field and passes the polodial-pointing signal fields sourced by axion conversion into photons. 
With this arrangement, axion re-conversion will result in a non-zero AC magnetic field in the center of the toroid. 
This field may be detected by a pick-up loop and amplified by coupling into the SRF receiver cavity. 

\begin{figure} [t]
   \begin{center}
   \begin{tabular}{c}
   \includegraphics[width=\columnwidth]{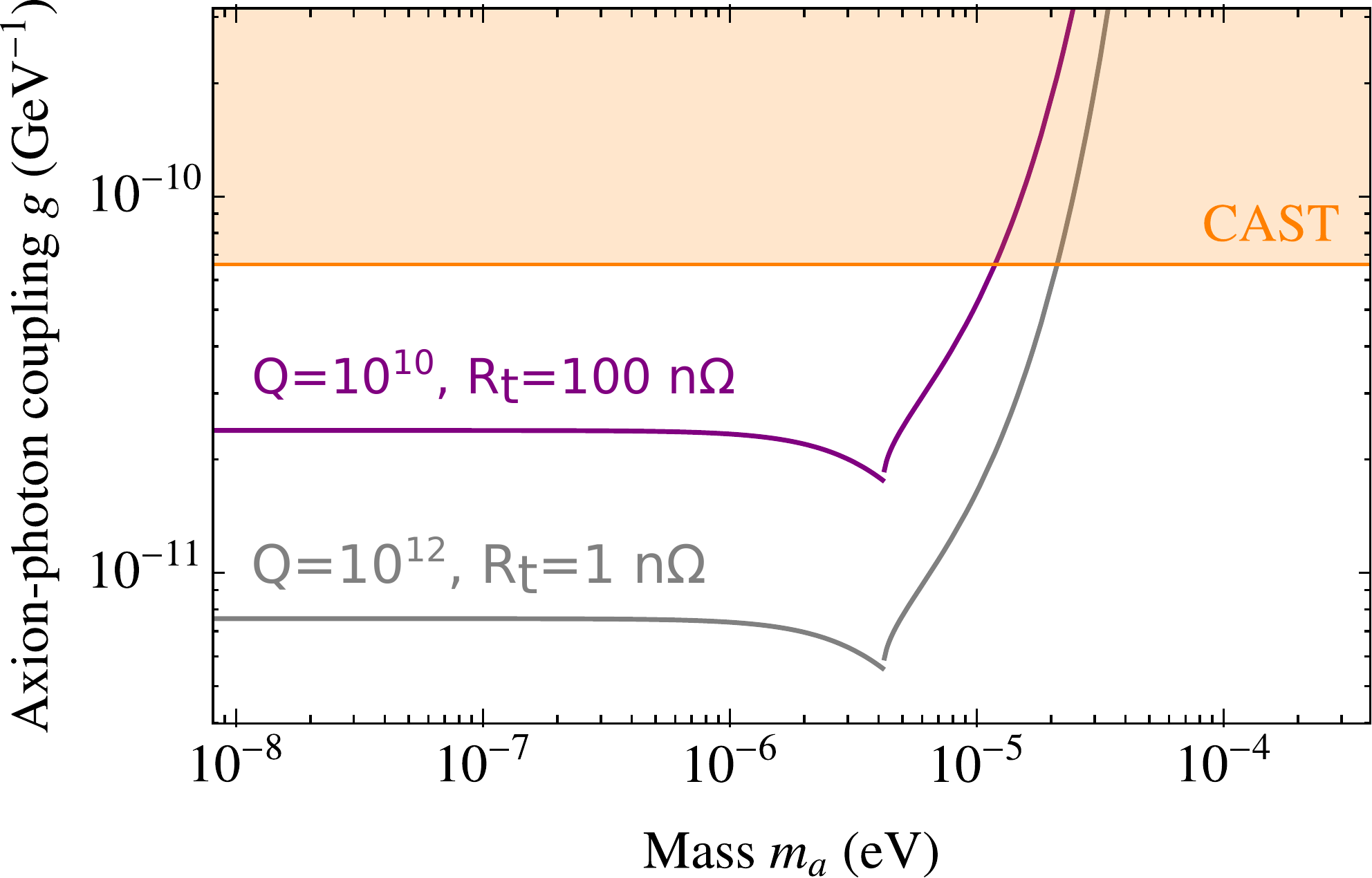}
   \end{tabular}
   \end{center}
   \caption 
   { \label{fig:LSWsensitivity} 
   Sensitivity to axion-photon coupling $g_{a\gamma\gamma}$ of the LSW search~\cite{Janish:2019dpr}.   See text for details.  The yellow region is constrained by solar axion detection~\cite{Anastassopoulos:2017ftl}. 
   }
   \end{figure}

With this approach neither SRF cavity is subject to large magnetic fields and neither suffers a degradation of $Q$-factor. However, losses in the walls of the conversion region can result in a decrease of the effective $Q$ of the entire system.  
The SNR is given by 
\begin{equation}
    \text{SNR} \sim \frac{g^4 B_c^4 B_0^2 L_t}{\omega^5} \,  \beta^2  \, \text{Min}\left[Q, \frac{\omega L_t}{R_t} \right] \, \frac{t_\text{int}}{T_\text{sys}}
    ~.
\end{equation}
$T_\text{sys}$ is the system temperature, with noise dominated by either thermal or quantum fluctuations.
$B_0$ is the static magnetic field in the conversion region, and $B_c \approx 0.2 \, \text{T}$ the SRF cavity's lower critical field which limits the power pumped into the source cavity. 
The signal power scales linearly with integration time $t_\text{int}$ as the phase of the input signal is known. 
$\beta \sim 10^{-3}$ is a geometric factor, accounting for the propagation of the signal between the production, conversion, and detection regions.  
$R_t$ is the toroid's surface resistance and $L_t$ its inductance for toroidal currents. 
For sufficiently small $L_t/R_t$, the system's $Q$ is set by the toroid's properties.
The toroid surface resistance is estimated in~\cite{Janish:2019dpr} to be $R_t \sim 100 \, \text{n}\Omega$, due to the fringing components of the static conversion field which will lead to flux vorticies in the toroid perpendicular to the flow of signal current. 
This resistance is limiting for $Q \geq 10^{10}$. 
To take full advantage of a cavity with $Q\sim 10^{12}$, the fringe fields must be sufficiently suppressed to allow $R_t \sim \text{n}\Omega$. 
The reach in both cases is given in Fig.\ref{fig:LSWsensitivity}, assuming a cavity with source frequency $\nu = \text{GHz}$ and integration time $t_\text{int} = 30 \, \text{days}$, taking $B_0 = 5 \ \text{T}$ and $T_\text{sys} = 0.1\ \text{K}$. 
$L_t$ is proportional to the toroid radius and is taken to be $L_t = 125 \, \text{nH}$ which corresponds to a $10\, \text{cm}$ toroid. 
Reaching these sensitivities will require exquisite control over the receiver cavity mode frequencies, as it is essential that a resonant mode of this cavity lies within $Q^{-1}$ of the source frequency. This will likely require continual active control to keep the receiver cavity tuned to the source.

\subsubsection{Two Cavities with a pump mode}

\begin{figure}[t]
\begin{center}
\includegraphics[width=\columnwidth]{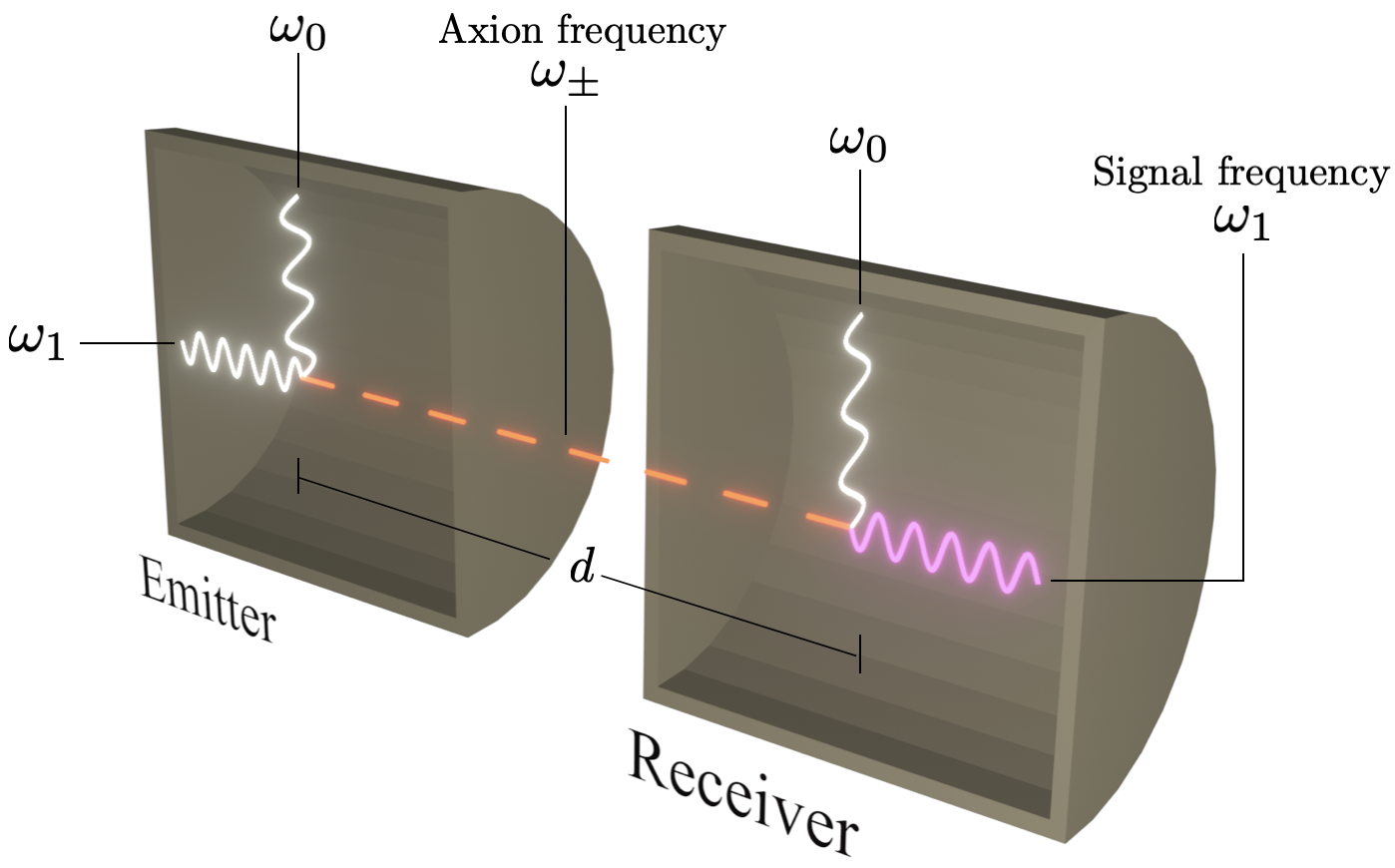}
\caption{Setup for axion search using two cavities separated by a distance $d$, with a pump mode with a frequency $\omega_0$ in the receiver and two active modes $\omega_{0,1}$ in the emitter. 
Axion is produced from the emitter with a frequency equal to the sum or difference of the two active modes, $\omega_{\pm}=\omega_1\pm\omega_0$. The axion may in turn produce a photon in the receiver with frequencies equal to $\omega_{\pm}\pm\omega_0$, in which the combination that yields $\omega_1$ can be resonantly produced.}
\label{fig:2cavity3modes}
\end{center}
\end{figure}
The LSW proposal above introduces a conversion region in between the production and detection cavities, so a large static $B$-field can be run there and turn the produced axions into signal photons. 
An alternative approach as shown schematically in Fig.~\ref{fig:2cavity3modes}, is to replace the static $B$-field with an oscillatory $B$-field, which can then be directly run inside the receiver cavity. The oscillatory $B$-field naturally comes about as a mode of the cavity, which is the pump mode.
Given an integration time $t_{\rm int}$, the SNR is approximately given by Equation~(\ref{eq:SNR}).
with the signal bandwidth $\Delta\omega_1$ is traditionally chosen as $\omega_1/Q$ but can be as small as $t^{-1}_{\rm int}$~\cite{Janish:2019dpr}.
We have assumed that the dominant noise is from thermal noise in Eq.~(\ref{eq:SNR}), characterized by the temperature $T$.
Other sources of noise will be discussed later.
Since the axion signal frequency is $\omega_{a}=\omega_1\mp\omega_0$, we have that $\omega_{a}\sim \omega_1\sim\omega_0\sim V^{-1/3}$, where $V$ is the volume of the cavity. 
In the limit that the axion is massless, taking $\Delta\omega_1$ to be $t_{\rm int}^{-1}$ yields
\begin{equation}
\begin{split}
\mbox{SNR}\sim& \frac{Q}{8\omega_1}V g_{a\gamma \gamma}^2
E^2_{\rm peak}
\left(\omega_{a}\frac{\eta^2_{01} g_{a\gamma \gamma}V E^2_{\rm peak}}{4\pi d}
\right)^2  \frac 1T t_{\rm int}\\
\sim &
5\,
\left(\frac{Q}{10^{10}}\right)
\left(\frac{V}{(0.2\rm m)^3}\right)^3 
\left(\frac {g_{a\gamma \gamma}\, \rm GeV }{5\times10^{-11} }\right)^4 \left(\frac{E_{\rm peak}}{80\rm MV/m}\right)^6\\
&\quad\times
\left(\frac{ 0.4\rm m}d\right)^2 \left(\frac{ \omega_a}{ \rm GHz}\right)\left(\frac { t_{\rm int}} {1\rm year}\right)\left(\frac {1.4\rm K} T\right)
\end{split}
\end{equation}
where both the emitter and receiver cavities are to have the same volume $V$ and the 3 active modes are assumed to have the same peak field value $E_{\rm peak}$.

The SNR above assumed that the only important background is from thermal fluctuations. In order to reach this level of sensitivity one must mitigate other sources of background. 
A particular worry of this multi-mode setup is that the signal mode lives in the same cavity as a spectator mode which is being driven to high occupancy. A small leakage of power either from the driving source or from the spectator mode to the signal mode can easily dominate over the thermal background. 
To put this in perspective, an excited mode with $E_\mathrm{peak}\sim 80$~MV/m at GHz frequencies has roughly $10^{26}$  photons. The thermal background at these frequencies is of order a few thousand photons. 
These challenges were already identified in Ref.~\cite{Sikivie:2010fa}. In Ref.~\cite{Gao:2020anb}, it is shown that both of these noise sources can be mitigated by using a pump with high $Q$ and with the pump frequency well separated from the signal mode frequency. In addition, such noise sources can be further suppressed by optimizing the cavity geometry and material science techniques to reduce nonlinearities. As opposed to DM searches, the signal in a LSW experiment may be turned off by deactivating or detuning the emitter in order to characterize the noise in the receiver cavity.

\subsubsection{Single-Cavity Axion Search and Euler-Heisenberg}

  \begin{figure} [t]
   \begin{center}
   \begin{tabular}{c}
   \includegraphics[width=\columnwidth]{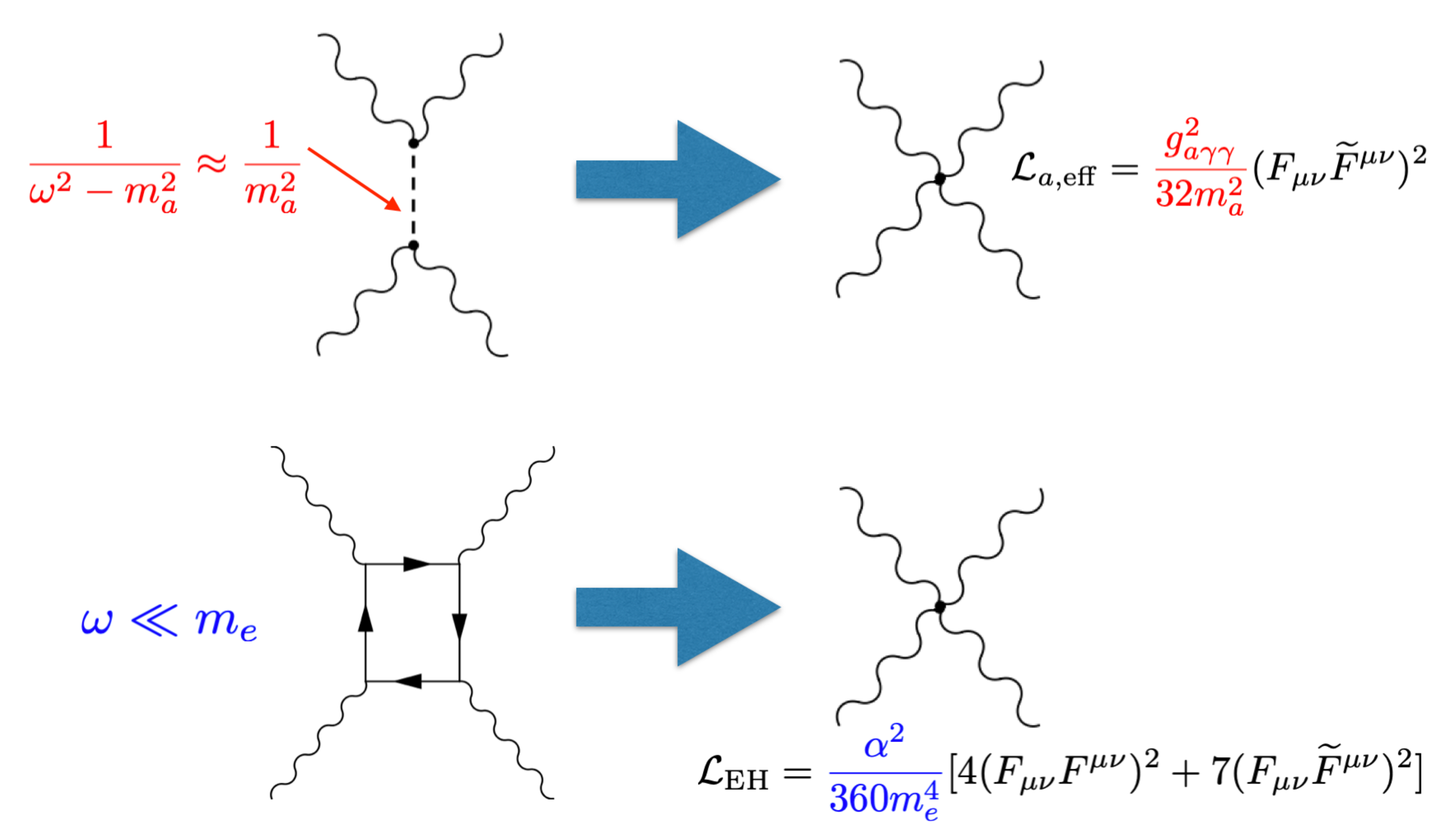}
   \end{tabular}
   \end{center}
   \caption[example] 
   { \label{fig:Leff} 
Integrating out an axion leads to an effective quartic nonlinearity in the QED Lagrangian (top). Integrating out the electron leads to a similar nonlinear Lagrangian, the Euler-Heisenberg Lagrangian, that is present in pure QED (bottom).}
   \end{figure} 
   
The presence of an axion with a coupling $\g$ to electromagnetic fields induces small deviations from the linearity of electrodynamics~\cite{Bernard:1997kj,Evans:2018qwy}, and axion searches can thus also be understood as precision tests of quantum electrodynamics (QED). Famously, loop contributions from virtual electrons will also induce such nonlinearities in pure QED, which are parameterized by the Euler-Heisenberg~(EH) Lagrangian~\cite{Heisenberg:1935qt,Schwinger:1951nm}. The EH Lagrangian makes a prediction for light-by-light scattering within the SM, which has never been observed at photon frequencies below the electron mass $m_e = 511 \ {\rm keV}$ because the effect is highly suppressed at low energies. Here we will describe a proposed experiment to search for both the axion-induced and EH nonlinearities using high-$Q$ SRF cavities. The operating principle~\cite{Brodin:2001zz,Eriksson:2004cz,Bogorad:2019pbu} is that in an SRF cavity simultaneously pumped at two resonant frequencies $\omega_1$ and $\omega_2$, the enormous occupation numbers of the pumped modes could generate a weak but observable signal at the signal frequency $ \omega_s = 2\omega_1 - \omega_2$ if the cavity geometry is tuned to make $\omega_s$ a resonant mode as well.

Consider the Feynman diagram in Fig.~\ref{fig:Leff}, top left, which represents photon-photon scattering mediated by an intermediate axion. When $m_a$ is large compared to the energies of the scattering photons, the axion may be ``integrated out'' by replacing its propagator as $(\omega^2 - m_a^2)^{-1} \to - m_a^{-2}$. This leads to a local Lagrangian of the form shown in Fig.~\ref{fig:Leff}, top right, which describes a 4-photon interaction, or equivalently a cubic nonlinearity in the equations of motion.

This argument has a close analogy with the EH Lagrangian, which is the low-energy limit of QED when the electron has been integrated out, $\omega \ll m_e$. Feynman diagrams of the type shown in Fig.~\ref{fig:Leff}, bottom left, give rise to a local interaction with the Lagrangian given in Fig.~\ref{fig:Leff}, bottom right. The nonlinearities of the EH Lagrangian may be described in terms of an effective charge and current density
\begin{equation}
\label{eq:JEH}
\rho_{\rm EH} \!=\! - \frac{4\alpha^2}{45m_e^4} \nabla \cdot \Pol \, ; \ 
	\J_{\rm EH} \!=\! \frac{4\alpha^2}{45m_e^4} \! \left( \nabla\! \times\! \M \!+\! \frac{\partial \Pol}{dt} \right)  ,
\end{equation}
with $\Pol = 7(\E\cdot \B)\B + 2 (\E^2 - \B^2)\E \,$ and $\M = 7(\E\cdot \B)\E - 2 (\E^2 - \B^2)\B\,$. Note that both $\Pol$ and $\M$ are cubic in the electromagnetic fields, and will thus source cubic nonlinearities; for example, an oscillating $E$-field at frequency $\omega$ will source a response field at frequency $3\omega$.

The axion-induced nonlinearities may also be \emph{non-local} when the axion is light, $\omega \gg m_a$, in which case a simple Lagrangian description is not available. The situation for any $m_a$ may still be parameterized in terms of an effective current,
\begin{equation}
\label{eq:Ja}
\rho_a = -\g \B \cdot \nabla a \, ; \ 
	\J_a = \g \left( \nabla a \times \E + \B \frac{\partial a}{\partial t} \right),
\end{equation}
where $a$ is the intermediate axion field determined by integrating a Green's function. As we will see below, the form of $a$ will depend on the relative sizes of $m_a$ and the frequencies of the external $E$- and $B$-fields, but as with the EH Lagrangian, $\rho_a$ and $\J_a$ are cubic in the electromagnetic fields for any $m_a$. 

A quick estimate may be made for the size of $\g$ at which axion-induced nonlinearities will be of the same order of magnitude as the EH nonlinearities. Comparing the coefficients of the two local Lagrangians in Fig.~\ref{fig:Leff}, we have that the axion-induced effect dominates for
\begin{equation}
    \label{eq:EHCompare}
	\frac{\g}{m_a} 
	\gtrsim 
	\mathcal{O}(1) \times \frac{\alpha}{m_e^2} 
	\simeq 
	\frac{10^{-10} \ \mathrm{GeV}^{-1}}{10^{-6} \ \mathrm{eV}}
	~.
\end{equation}
Since $10^{-6} \ \mathrm{eV}$ corresponds to GHz frequencies, and the strongest limits on $\g$ which are agnostic to whether or not the axion makes up the cosmological DM of the universe are $\g < 6.6 \times 10^{-11} \, \mathrm{GeV}^{-1}$ from the CAST experiment~\cite{Anastassopoulos:2017ftl}, Eq.~(\ref{eq:EHCompare}) shows that an experiment sensitive to the EH nonlinearities could also set the strongest bounds on the axion-photon coupling.

  \begin{figure}[t]
   \begin{center}
   \begin{tabular}{c}
   \includegraphics[width=\columnwidth]{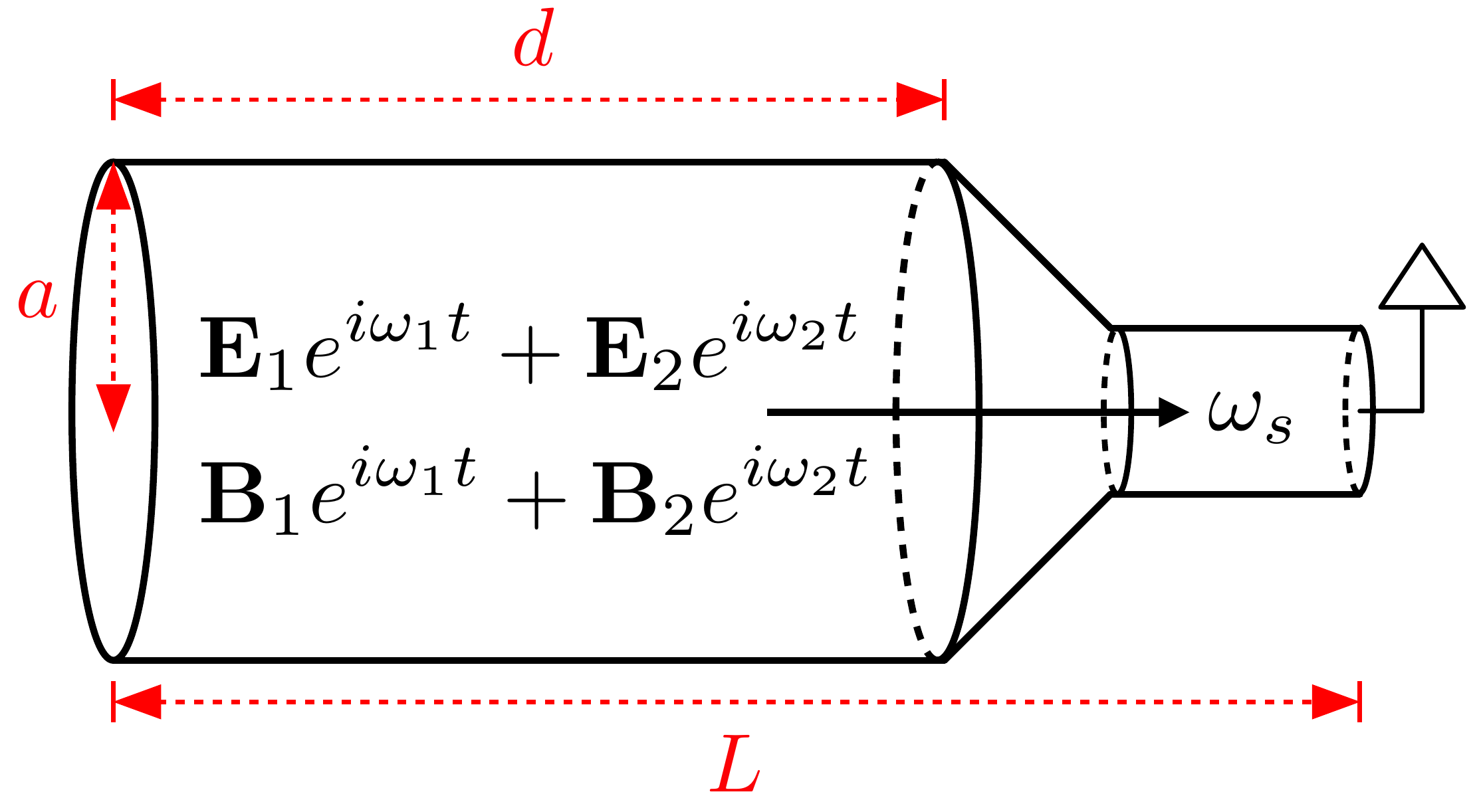}
   \end{tabular}
   \end{center}
   \caption[example] 
   { \label{fig:Schematic} 
Schematic of the proposed single-cavity experiment~\cite{Eriksson:2004cz,Bogorad:2019pbu}. Note that the signal mode $\omega_s$ extends through the bulk of the cavity, but the filtering geometry suppresses the pump fields in the detection region.}
   \end{figure}

Consider an SRF cavity pumped simultaneously with resonant electromagnetic fields $\E_1, \B_1$ and $\E_2, \B_2$ with frequencies $\omega_1$ and $\omega_2$ respectively, as illustrated in Fig.~\ref{fig:Schematic}. The intermediate axion field can be obtained by integrating the source $(\E_1 + \B_1) \cdot (\E_2 + \B_2)$ against the massive Klein-Gordon Green's function, yielding an axion effective current from Eq.~(\ref{eq:Ja}) containing a frequency component at $\omega_s = 2 \omega_1 - \omega_2$,
\begin{equation}
    \label{eq:j-cos-sin}
    \J_a(x, t) \supset \J_a^{(c)}(x) \cos(\omega_s t) + \J_a^{(s)}(x) \sin(\omega_s t)
    ~.
\end{equation}
For general $m_a$, both quadratures in $\J_a$ are present, but in the local limit $m_a \gg \omega_s$, $\J_a$ is always in phase with the pump modes. Let $\hat{\E}_s$ be the dimensionless signal mode in a cavity with volume $V$, with normalization $\int d^3 x\, |\hat{\E}_{s}(x)|^2 = V$, and quality factor $Q_s$. The axion-sourced signal field $\E_a$ is then~\cite{Bogorad:2019pbu}
\begin{equation}
	\label{eq:ESolFull}
	\E_{a} (x,t) 
= 	\frac{Q_s}{\omega_s V}\hat{\E}_s(x)\int d^3 x' \, \hat{\E}_s(x') \cdot \J_a(x',t)~, \end{equation}
from which we can obtain the total number of photons in the signal field as
\begin{equation}
	\label{eq:N3}
	N_s 
= 	\frac{1}{2 \omega_s} \left \langle \int d^3 x \, |\E_a(x,t)|^2 \right \rangle,
\end{equation}
where the brackets denote time averaging. Note that all of the above considerations hold equally well for the EH-sourced signal field, using Eq.~(\ref{eq:JEH}) instead for the effective current.

\begin{figure}[t]
   \begin{center}
   \begin{tabular}{c}
   \includegraphics[width=0.95\columnwidth]{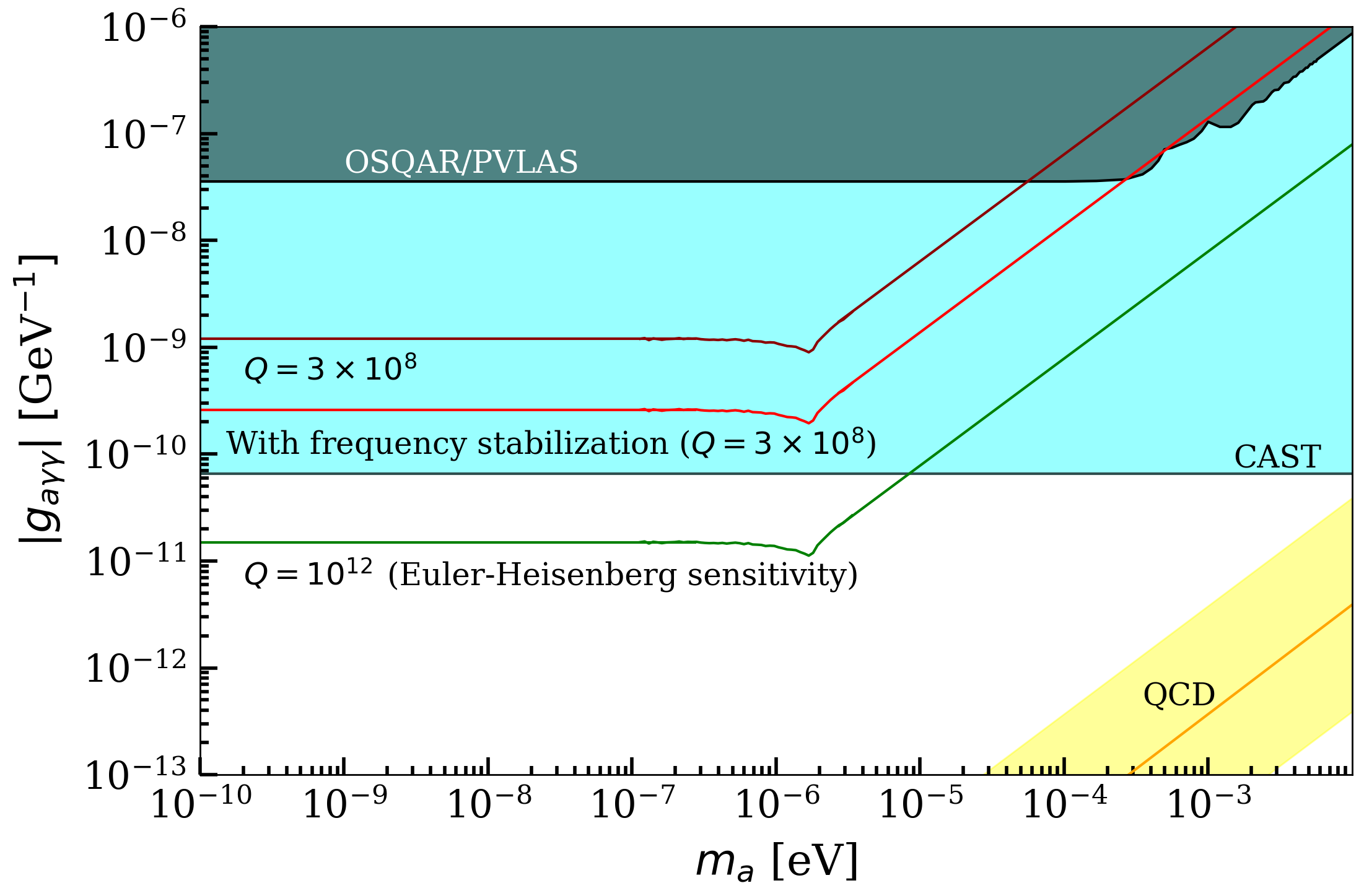}
   \end{tabular}
   \end{center}
   \caption[example] 
   { \label{fig:Sensitivity} 
Projected sensitivity of the single-cavity experiment with different assumptions about the cavity parameters, bandwidth, and measurement time; see text for details.}
   \end{figure} 

We can obtain an estimate of the sensitivity by assuming that thermal noise dominates the background and using the Dicke radiometer equation for the SNR,
\begin{equation}
	{\rm SNR} = \frac{P_s}{T} \sqrt{\frac{t}{B}} \approx \frac{N_s}{N_{\rm th}} \frac{1}{2 L Q_s} \sqrt{\frac{t}{B}},
\end{equation}
where $P_s$ is the signal power, $t$ is the total measurement time, $B$ is the signal bandwidth, $L$ is the length of the cavity, and $N_{\rm th} = T/\omega_s$ is the number of thermal photons at the signal frequency (valid for temperatures $T \gg \omega_s$). Setting SNR = 1 and solving for $\g$ for a cylindrical cavity with $a = 0.5$~m and $d = 1.56$~m, temperature $T = 1.5$~K, peak pump field strength $E_0 = 45$~MV/m, and the mode combination TE$_{011}$/TM$_{010}$/TM$_{020}$ (with $f_s  = \omega_s/(2\pi) = 0.527 \ {\rm GHz}$) yields the projected sensitivities shown in Fig.~\ref{fig:Schematic}, for various measurement times and readout schemes. The red curves correspond to $Q_s = 3 \times 10^{8}$, a measurement time of $t = 1$~day, and a bandwidth $B = \frac{\omega_s}{2\pi Q_s} = 2$~Hz (dark red) or $B = 1/t$ (light red); the latter may be achieved with a phase-sensitive readout such as a lock-in amplifier, because the phase of the signal is known according to Eq.~(\ref{eq:j-cos-sin}). The green curve corresponds to $Q_s = 10^{12}$, $t = 20$~days, and $B = 1/t$, and would achieve sensitivity to the EH nonlinearity. In order for $B = 1/t$ to be viable, the frequencies of the cavity, and in particular the mode-matching condition, must be robust to vibrational noise, possibly with an active frequency-stabilizing scheme. 

For all of the sensitivity curves, the limiting sensitivity to $\g$ is independent of $m_a$ for $m_a \ll \omega_s$ and degrades linearly with $m_a$ for $m_a \gg \omega_s$. This is one potential advantage of the single-cavity scheme compared to the two-cavity schemes: since the signal generation in the single-cavity scheme takes place in the same spacetime region as the pump region, a heavy off-shell axion does not have to propagate to be converted back to photons, a fact which causes the reach of the two-cavity scheme to degrade exponentially at large $m_a$. On the other hand, the single-cavity scheme is far more sensitive to noise sources which generate nonlinearities in the pump region, so these two schemes are highly complementary.


\subsubsection{Ongoing R\&D for Axion searches}

The previously discussed axion searches will require custom designed SRF cavities in order to fulfill the stringent requirements on frequency matching between the pump mode(s) and the signal mode. In addition, multiple issues, both technical and scientific, need to be addressed in order to maximize the search sensitivity. A feasibility study is being conducted to evaluate how the multimode axion searches can be implemented with the necessary sensitivity.

An elliptical 1.3~GHz  9-cell SRF Nb cavity is being used to study the presence and magnitude of non-linear effects of the Nb surface that could cause the excitation of frequencies or resonant modes other than the ones intentionally excited in the cavity. The geometry of this cavity, being multi-cell, is different from the single cell cavities that will be used in the axion searches. However, the multi-cell design allows for easy access to multiple resonant modes, both nearby ($ \approx$MHz apart) or well-separated ($\approx$GHz) in frequency. The measurements are being conducted at 2 K in a liquid helium dewar and include:
\begin{itemize}
    \item the excitation of one pump mode and the search for excited modes both in the same resonant pass-band or in different TM or TE pass-bands.
    \item the excitation of two pump modes and the search for their linear combinations, focusing in particular on the combinations that would correspond to the axion signal: $\omega_{3}=2 \omega_{1} \pm \omega_{2}$.
\end{itemize}
In addition to the study of non-linear effects, it is also crucial to identify the possible sources of spurious excitation in the RF system that could propagate to the cavity (as for example higher harmonics of the pump modes generated by the high power amplifier or resonant modes excited in the cavity by the amplifier noise). 
Analyzing the sources of noise in the current experimental system will help guide the choice of appropriate RF instruments and components to be used in the axion searches in order to maximize their sensitivity.

To guide the mechanical design of the axion cavities, effort is also being directed at characterizing the cavity mechanical modes and the microphonics present in the helium bath, and how these result in the cavity deformation and frequency shift for each resonant mode of interest. These developments will als directly benefit the development of the axion DM search discussed below in Sec.~\ref{sec:UpConversion}.

\section{Dark Matter Searches with Ultra-high Q}

Both axions and dark photons are compelling candidates for DM~\cite{Preskill:1982cy,Dine:1982ah}. These candidates fall into a broader category of wave-like DM, since their wave, rather than particle, nature provides a more useful description. As light bosonic degrees of freedom, axions or dark photons can form non-relativistic waves which would carry energy density, and hence gravitate. Focusing on the axion for concreteness, a non-relativistic wave is well-approximated as a coherent field oscillating with a frequency $\omega_a \simeq m_a$, where $m_a$ is the axion mass. 
The amplitude, $a_0$ of the axion field oscillations is set by the DM energy density $\rhodm$ via $a_0\sim\sqrt{\rhodm}/m_a$.
The DM axion wave is expected to carry momentum of roughly $m_a v$ with $v \sim 10^{-3} \ c$ of order the local virial velocity in our halo. This leads to a small average correction to the axion oscillation frequency, of order $v^2$. This implies that the effective quality factor of the axion as a monochromatic source is $Q_a\sim 10^6$. Hence, the axion field is expected to be coherent over roughly a million periods, by which time the axion picks up a random phase with respect to some initial state. In addition, the axion momentum implies that there is a small $\order{v}$ gradient in the axion field, though in many cases it can be ignored. A similar story follows for dark photon DM, with the notable exception that the dark photon is a vector field, and hence picks a random spatial direction which changes over a coherence time.

\begin{table}[t]
\begin{tabular}{ c l c c c c}
\hline\hline
 Ref. & Material & $f$ (GHz) & $B_a$ (T) & $T$ (K) & $Q_0$ \\ \hline 
 \cite{Alesini:2019ajt} & Nb$_3$Sn & 3.9 & 6.0 & 4.2 & $(5.3\pm0.3)\times10^5$ \\  
 \cite{Golm:2021ooj} & NbTi/Cu & 9.08 & 5 & 4.2 & $2.95\times10^5$ \\  
 \cite{Ahn:2019nfy} & Nb$_3$Sn & 9 & 8 & 4.2 & $6\times10^3$ \\  
 \cite{divora:2022} & REBCO & 9 & 11.6 & 4.2 & $7\times10^4$ \\  
 \cite{DiVora:2022tro} & YBCO & 6.93 & 8.0 & 4.2 & $3.2\times10^5$\\
 \hline\hline
\end{tabular}
\caption{A list of existing cavities which were operated in strong magnetic fields and the corresponding quality factors.}
   \label{tab:Bcavitytable} 
\end{table}

It should be noted that both axions and dark photons are attractive candidates also because the abundance of DM can be explained by a random initial displacement of the field. For the axion this can occur either during the QCD phase transition or during inflation~\cite{Dine:1982ah,Preskill:1982cy}. For the QCD axion, an initial displacement of order the axion decay constant motivates searches around a GHz, but lower frequencies can also arise naturally.  
Several mechanisms could produce cosmic dark photons, the simplest being through quantum fluctuations during inflation~\cite{PhysRevD.93.103520}. These fluctuations seed excitations in the dark photon field, resulting in the cold DM observed today in the form of coherent oscillation of this field. The predicted mass from this mechanism is 
$m_{A^{\prime}}\approx 10\,{\mu \mathrm{eV}} \left( 10^{14}\,\mathrm{GeV} /H_I\right)^4$, 
where $H_I$ is the Hubble constant during inflation. Measurements of the CMB tensor to scalar ratio constrain 
$H_I \lesssim 10^{14}$~GeV~\cite{Planck:2015sxf}, 
which makes the search for $m_{A^{\prime}}\gtrsim 10^{-5}$~eV well-motivated. Other mechanisms are possible and are described in Refs.~\cite{PhysRevD.104.095029, Arias_2012}.

\subsection{Haloscopes and the Scan Rate}

A scheme that has proven successful to reach sensitivity relevant for QCD axion detection is that of the cavity haloscope. This detector relies on axion to photon conversion inside RF cavities in a strong magnetic field. Similarly a DM dark photon can convert to a photon, even in the absence of a magnetic field. The signal power read out of a haloscope with a cavity of volume $V$, a loaded quality $Q_l$, magnetic field $B$, and coupling $\beta$ is
\begin{equation}
    P_\mathrm{sig} = P_0
    \frac{\beta}{1+\beta}
    \frac{Q_l Q_\mathrm{DM}}{Q_l+Q_\mathrm{DM}}
\end{equation}
where 
\begin{equation}
    P_0 =\left\{
    \begin{array}{cl}
        g_{a\gamma\gamma}^2\frac{\rho_a}{m_a}B^2 V C_a 
        & \mbox{for axions} \\
        \epsilon m_{A'} \rho_{A'} V C_{A'} 
         & \mbox{for dark photons}
    \end{array} \right.
\end{equation}
and the $C$'s are form factors, normalized mode ovelaps between the DM field and the cavity mode, typically of order 0.1.

\begin{figure}[t]
   \centering
   \includegraphics[width=\columnwidth]{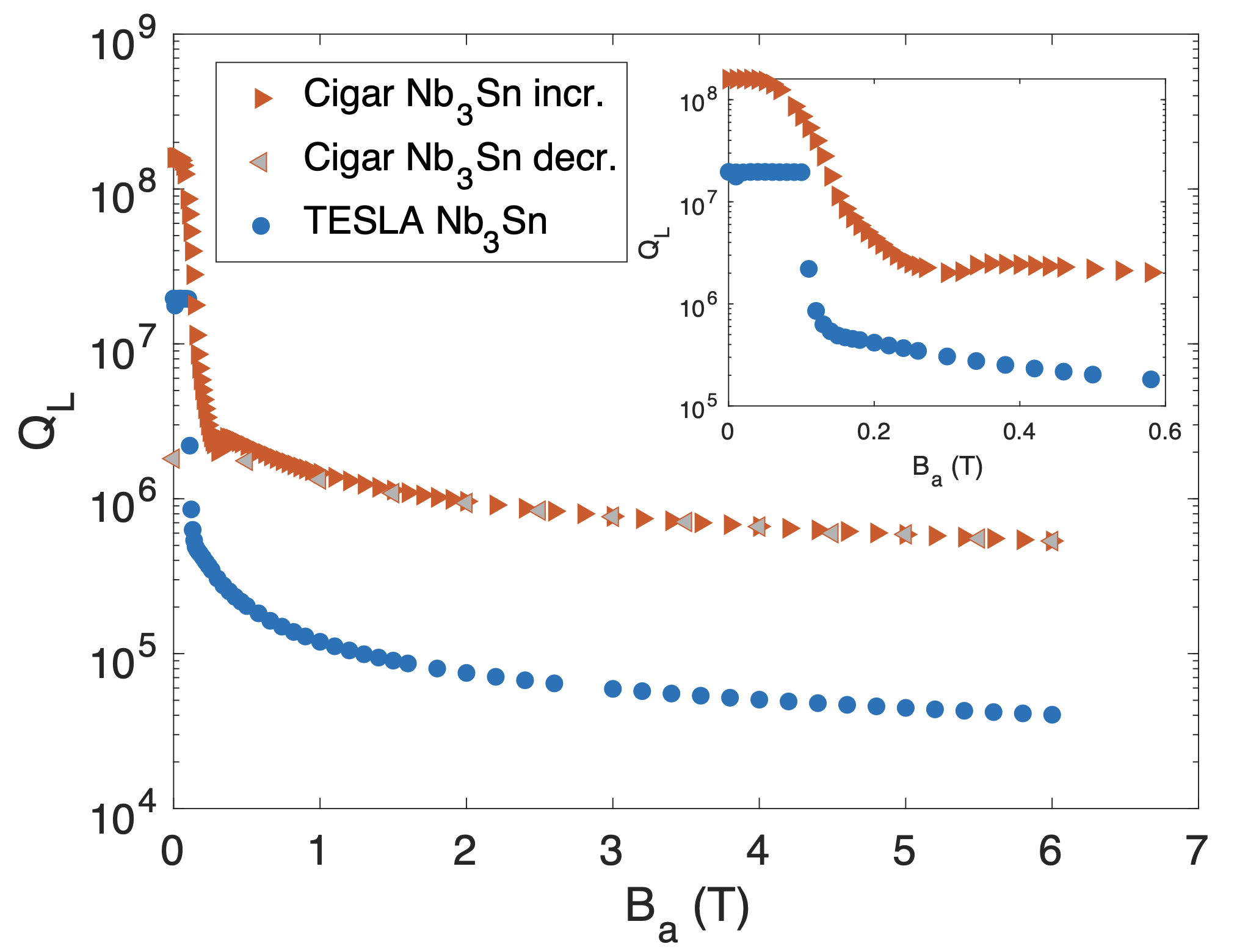}
   \caption{\label{fig:Nb3SnData} Loaded quality factor as a function of the applied magnetic field for a TESLA and a cigar-shaped Nb$_3$Sn cavity~\cite{Posen:2022tbs}. The inset focuses on the low-field region. For the cigar-shaped Nb$_3$Sn cavity, measurements were made both for increasing and for decreasing field.}
\end{figure}

The axion (dark photon) mass is unknown, so cavity haloscopes must be tunable to search through $g_{a\gamma \gamma}$ vs. $m_a$ ($\epsilon$ vs. $m_{\Ap}$) parameter space. Thus the scan rate is a key figure of merit for such experiments. Accounting for thermal and amplifier noise, it was shown that the scan rate allowed by fixing the SNR can be higher even as the cavity $Q$ exceeds the quality of the DM source~\cite{Kim_2020}. It is thus well motivated to use ultra-high quality cavities in the search for DM, and in the case of the axion, to develop high coherence cavities that can withstand large magnetic fields. 
 
\subsection{High-$Q$ Cavities in a Static $B$-field} 

\begin{figure}
    \centering
    \includegraphics[width=\linewidth]{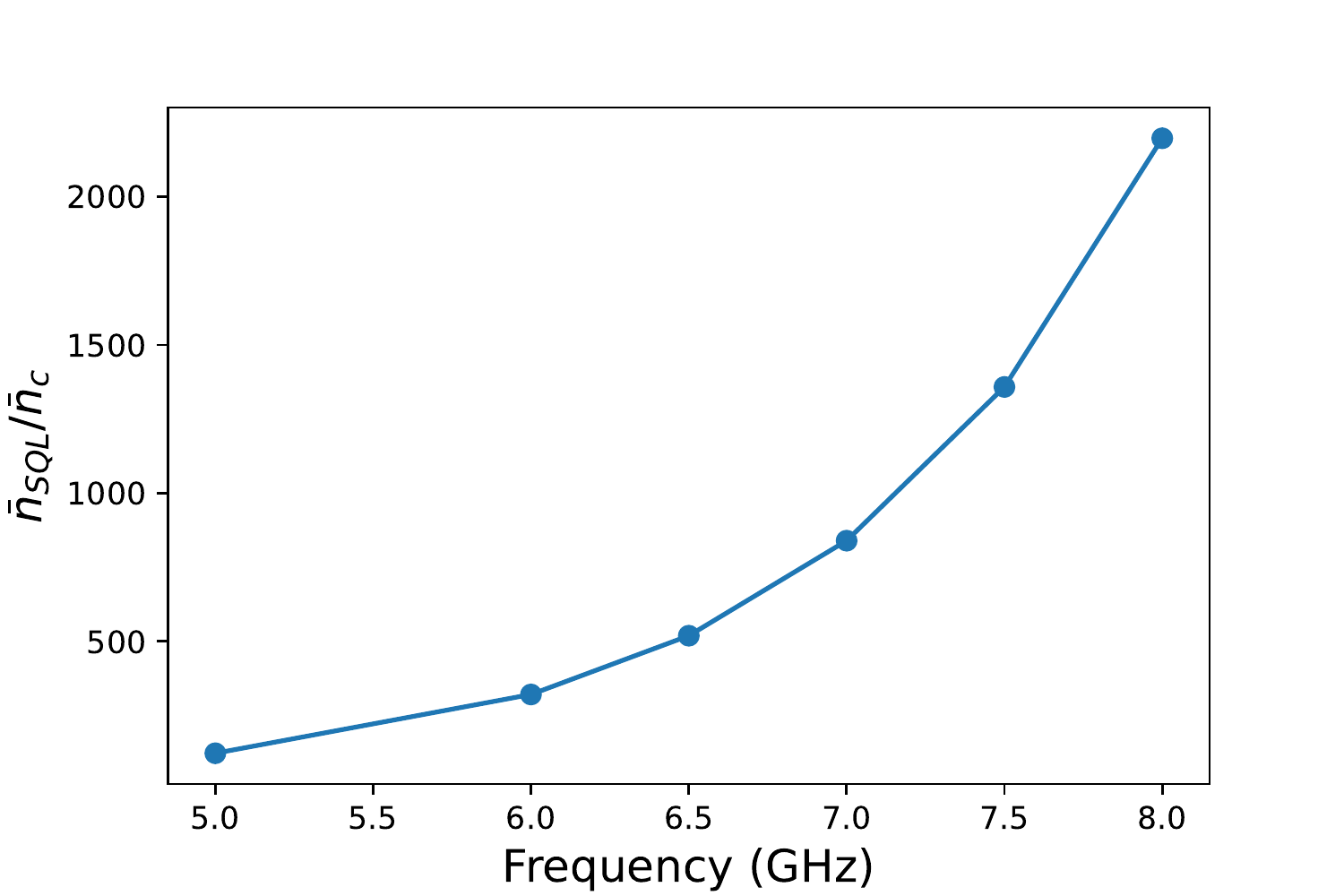}
    \caption{Implementing the photon counting technique will increase the scan rate by a factor $\bar{n}_{SQL}/\bar{n}_c$ compared to the standard quantum limit. The calculated $\bar{n}_c$ assumes a 50~mK cavity and that only real thermal photons contribute to the dark count.}
    \label{fig:dpdm_dfdt}
\end{figure}

Using large cylindrical copper cavities, axions with mass in the range $(2.7-4.2) \ \mu \text{eV}$ (650\,MHz $-$ 1025\,MHz) have been excluded at DFSZ model sensitivity~\cite{ADMX}. To extend the search to heavier axions (up to 15\,GHz), next generation experiments target the development of new cavity technologies with higher quality factors to compensate for the poor scaling of the scan rate with increasing frequency, mostly determined by the effective volume of the chosen axion-sensitive cavity mode. For copper cavities employed so far in leading haloscope experiments, the quality factor is limited by the anomalous skin effect~\cite{skin:depth} to $\sim 10^4-10^5$ at $\sim$\,GHz frequencies~\cite{Alesini:2021}. Thus, radically different approaches need to be considered to realize better resonator haloscopes. 

\begin{figure*}[t]
    \centering
    \includegraphics[width=0.8\linewidth]{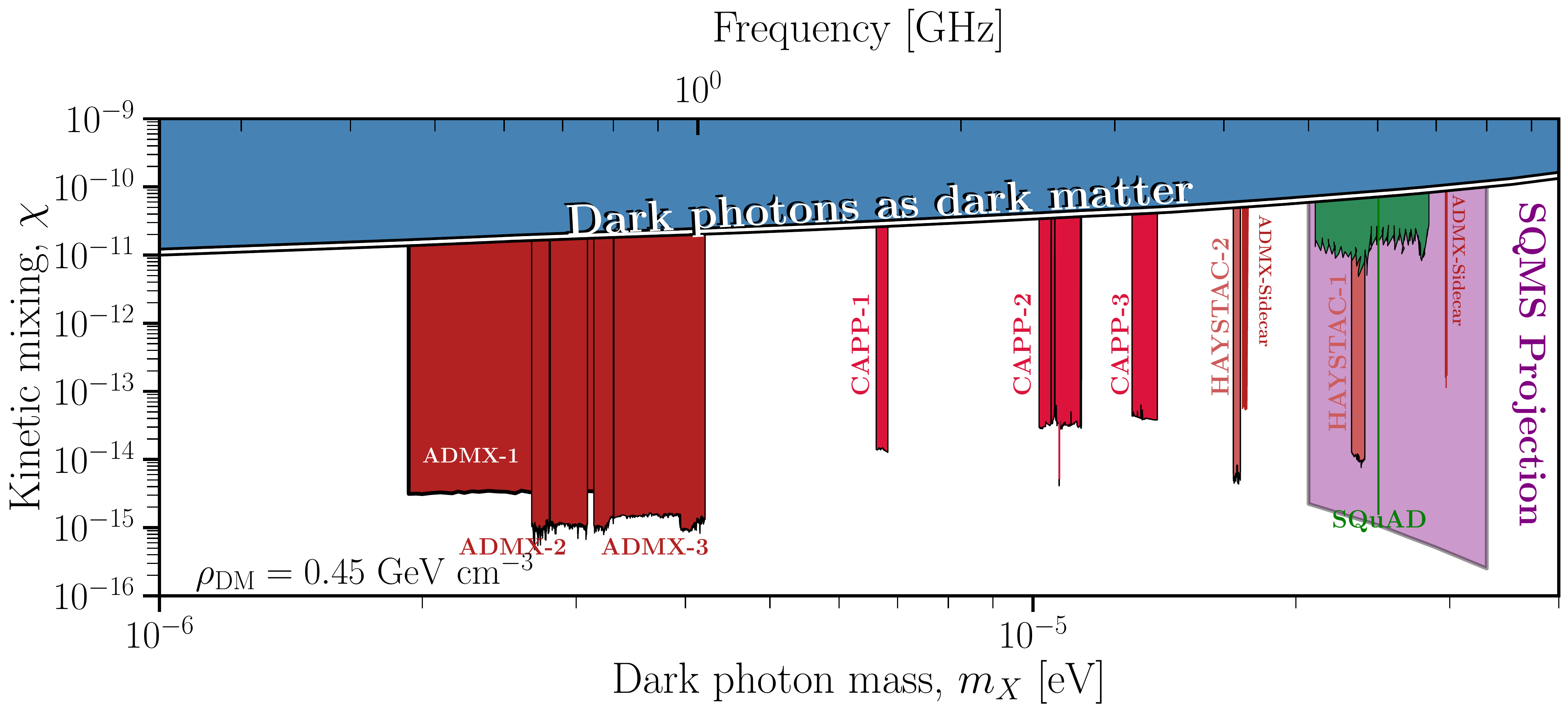}
    \caption{The projected sensitivity of the proposed search for dark photon dark matter with SQMS systems assuming a fixed scan rate.}
    \label{fig:dpdm_science_reach}
\end{figure*}

SRF cavities in accelerators typically are operated in small ${\sim}\mu T$-scale magnetic fields, with $Q_0 \sim \order{10^{10}}$. In these low magnetic field environments, niobium is expected to be in the Meissner state, in which the superconductor acts as a diamagnet. In $T$-scale magnetic fields, niobium would be normal conducting, but superconductors with higher critical fields like Nb$_3$Sn and NbTi can still be superconducting, though typically in the vortex state, in which applied magnetic fields penetrate through the material in a lattice of quantized flux. This trapped flux will degrade the $Q_0$ of the cavity to some extent, but the cavity may still have a substantially higher $Q_0$ than an equivalent normal conducting cavity.

Experiments exploring the performance of superconducting cavities in $T$-scale magnetic fields have been successful in producing $Q_0$ values in the mid-$10^5$ range at frequencies between 1-10 GHz, as shown in Table~\ref{tab:Bcavitytable}. These are promising for improving cavity haloscope-based axion search experiments, in which the sensitivity scales approximately with $\sqrt{Q_0}$ and the scan rate scales approximately with $Q_0$. The performance of a Nb$_3$Sn cavity from the SQMS collaboration is shown in Fig.~\ref{fig:Nb3SnData}. This cavity was measured only up to 6 T due to test stand limitations, but may continue to have high $Q_0$ up to even higher magnetic fields, and may have even higher $Q_0$ at lower temperatures than 4.2 K~\cite{Posen:2022tbs}.

Dielectric resonators~\cite{McAllister:2018, Kim:2019asb} are another effective strategy to obtain high quality factors in multi-Tesla fields, especially in the high-frequency range. 
In this type of cavity, higher-order TM$_{0n0}$ modes are exploited for detection, and the inserted dielectric shells or rings coaxial with the cavity axis are properly shaped to maximize the figure of merit $(C_{0n0}V)^2Q_0$, and in turn the haloscope scan rate, in FEM simulations. 
$C_{0n0}$ is the form factor, that accounts for the overlap between the wave functions of the axion and the produced photon. For cylindrical cavities, the TM$_{0n0}$ mode is conventionally considered since it yields the largest form factor in a magnetic field, while the higher-order modes have form factor values that decrease with increasing order. For instance, $C_{030}=0.05$ is parametrically smaller than  $C_{010}=0.69$ in an empty cavity. However, form factors that are very close to those of the fundamental mode can be in principle be obtained even for high-order modes if one dielectric shell is inserted in the cavity~\cite{McAllister:2018}.  
Alternatively, in the double-shell configuration, two concentric dielectric hollow cylinders are employed to suppress the evanescent field at the copper boundaries, to achieve quality factors sufficiently large to compensate the loss in form factor~\cite{divora:2022}.  The estimated scan rate of this latter dielectric resonator exceeds that of a conventional resonator by more than a factor 4, with measured quality factors exceeding $9\times 10^{6}$ at more than 8\,T magnetic fields.

\subsection{Sub-SQL Dark Photon Dark Matter Search with a Highly-Tunable SRF Cavities}

SQMS has recently started to design a dark photon dark matter (DPDM) search that will implement a widely tunable SRF cavity to scan the parameter space from 5~GHz to 8~GHz. SQMS will also take advantage of its in-house QIS expertise to reduce the noise floor below the Standard Quantum Limit (SQL) by using a superconducting qubit to count photons potentially deposited by a dark photon~\cite{PhysRevLett.126.141302}.



Cavity haloscopes have traditionally extracted the DM signal via an antenna connected to a linear amplifier, such as a Josephson Parametric Amplifier. Unfortunately, these linear amplifiers contribute to their own noise power, and their minimum contribution is the standard quantum limit (SQL). SQL noise increases linearly with frequency, and thus it is necessary to subvert the SQL to make higher-mass searches feasible. Two experiments have demonstrated sub-SQL detection by sacrificing phase information: HAYSTAC by implementing vacuum squeezing~\cite{Backes2021} and SQuAD by implementing qubit-based photon counting~\cite{Dixit:2020ymh}.

SQMS plans to increase the scan rate by implementing the photon-counting technique developed in Ref.~\cite{Dixit:2020ymh}. A transmon qubit will be situated inside the SRF cavity. The resonant frequency of the qubit will depend on the number of photons in the cavity, and Ramsey interferometry will be implemented to extract this resonant frequency. Since this is a quantum nondemolition measurement, this measurement can be repeated many times to reduce uncertainties and reject false positives. If the only source of false positives, i.e., dark count, comes from real thermal photons, then the scan rate would improve by a factor of $\mathcal{O}(10^2)$ to $\mathcal{O}(10^3)$ compared to SQL-limited detection. The photon occupation number for thermal photons in the cavity at temperature $T_c=50\,\mathrm{mK}$ is $\bar{n}_c=\big(\exp(hf / k_b T_c)-1\big)^{-1}$. The occupation number from the standard quantum limit is $\bar{n}_{SQL}$, and the scan rate improvement from subverting the SQL limit is $\bar{n}_{SQL}/\bar{n}_c$. The calculated scan rate improvement (assuming very generously that the dark count is dominated by the finite cavity temperature) is shown in Fig.~\ref{fig:dpdm_dfdt}.

The cavity conceptualized in this section will be sensitive to DPDM with masses between $20.3\,\mathrm{\mu eV}\, (5~\mathrm{GHz})$ and $33\,\mu\mathrm{eV}\, (8~\mathrm{GHz})$ with kinetic mixing angles of $\order{10^{-15}}-\order{10^{-16}}$, as shown in Fig.~\ref{fig:dpdm_science_reach}. The form factor varies from $C=0.05$ to $C=0.225$, assuming dark photons are randomly polarized.\footnote{Because dark photons are often thought to be randomly polarized, the form factor for a dark photon haloscope, in this case, is 1/3 that of an axion haloscope.} The Nb  cavity is expected to have an unloaded quality factor of $10^{10}$. However, the superconducting qubit is likely to reduce $Q_L$ by a few orders of magnitude. 

This project is currently at its conceptual and design stage. The projection is likely to fluctuate by an order of magnitude as this project matures. Regardless, SQMS is excited to combine SRF cavity technology and qubit-based photon counting to increase the DM search scan rate by many orders of magnitude.

  \begin{figure}[t!]
   \begin{center}
   \begin{tabular}{c}
   \includegraphics[width=0.5\columnwidth]{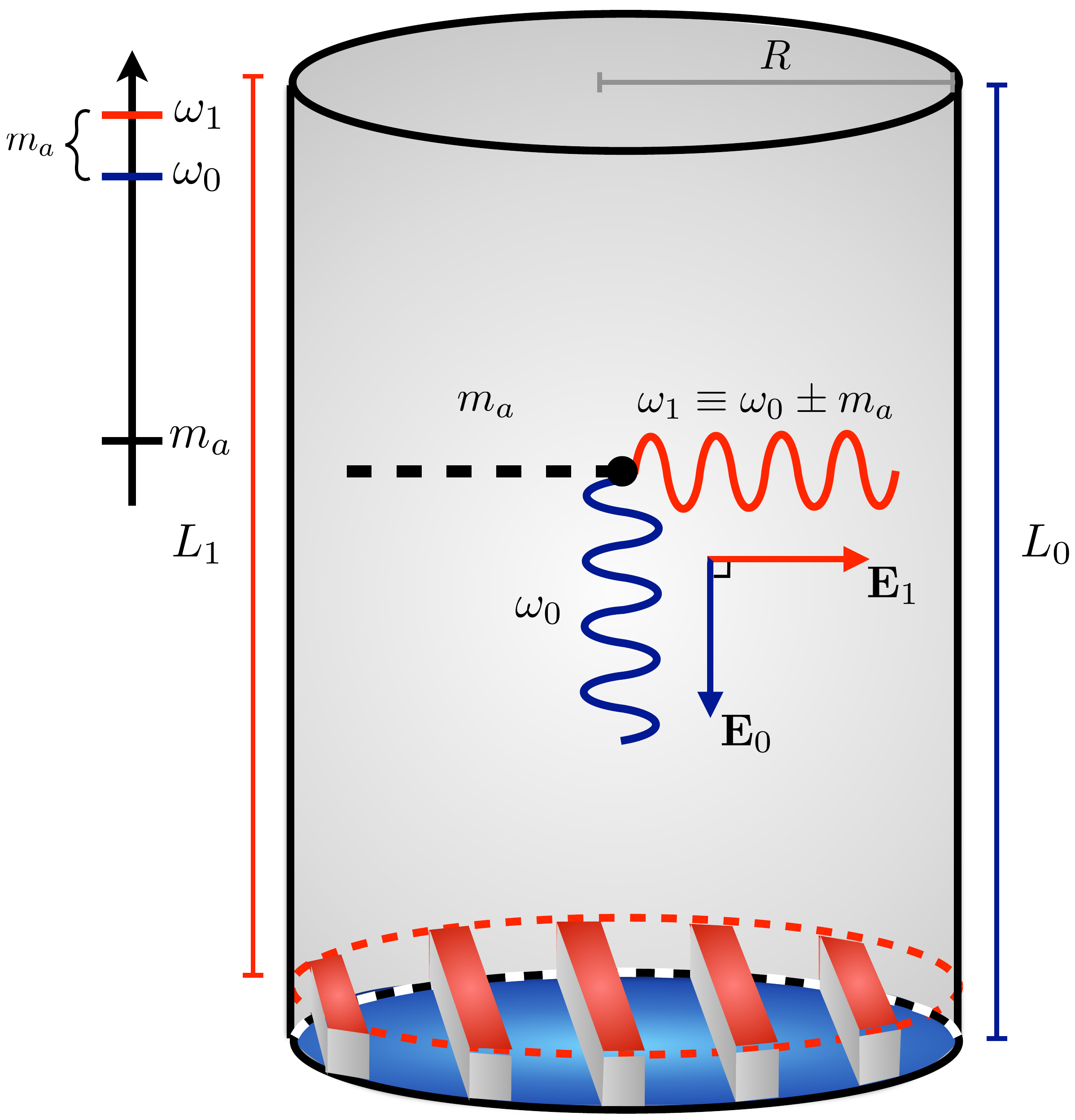}
   \end{tabular}
   \end{center}
   \caption[example] 
   { \label{fig:heterodynesetup} 
Schematic of the proposed up-conversion axion dark matter experiment. A pump mode photon at frequency $\omega_0 \sim \text{GHz}$ is upconverted by axion dark matter into a nearly degenerate photon of frequency $\omega_1 = \omega_0 + m_a$. See Ref.~\cite{Berlin:2019ahk} for additional details. 
}
   \end{figure}

\subsection{Axion Dark Matter Searches with Up-Conversion}
\label{sec:UpConversion}

To date, most experiments searching for electromagnetically-coupled axion DM employ \emph{static} magnetic fields, since these are more easily sourced at large field strengths. In the presence of an axion DM background, the axion-photon interaction sources a feeble electromagnetic field oscillating at the axion mass $m_a$, which can be detected with a resonator tuned to the same frequency. This is the strategy employed by typical haloscope cavity experiments, which operate at $\sim \text{GHz}$ frequencies, corresponding to axion masses of $m_a \sim \mu \text{eV}$. Since the resonant frequency is typically controlled by the inverse geometric size of the detector, probing much lighter axion masses requires prohibitively large cavities. To circumvent this, LC circuit resonator haloscopes are in the beginning stages of development to search for much lighter axions, enabled by the fact that their resonant frequency is not directly dictated by the size of the circuit~\cite{Chaudhuri:2014dla}. However, in the limit that the axion's Compton wavelength is much larger than the characteristic size of the detector $L_\text{det}$, the signal power is parametrically suppressed by the small quantity $m_a \, L_\text{det} \ll 1$.

  \begin{figure*}[t]
   \begin{center}
   \begin{tabular}{c}
   \includegraphics[width=1.5\columnwidth]{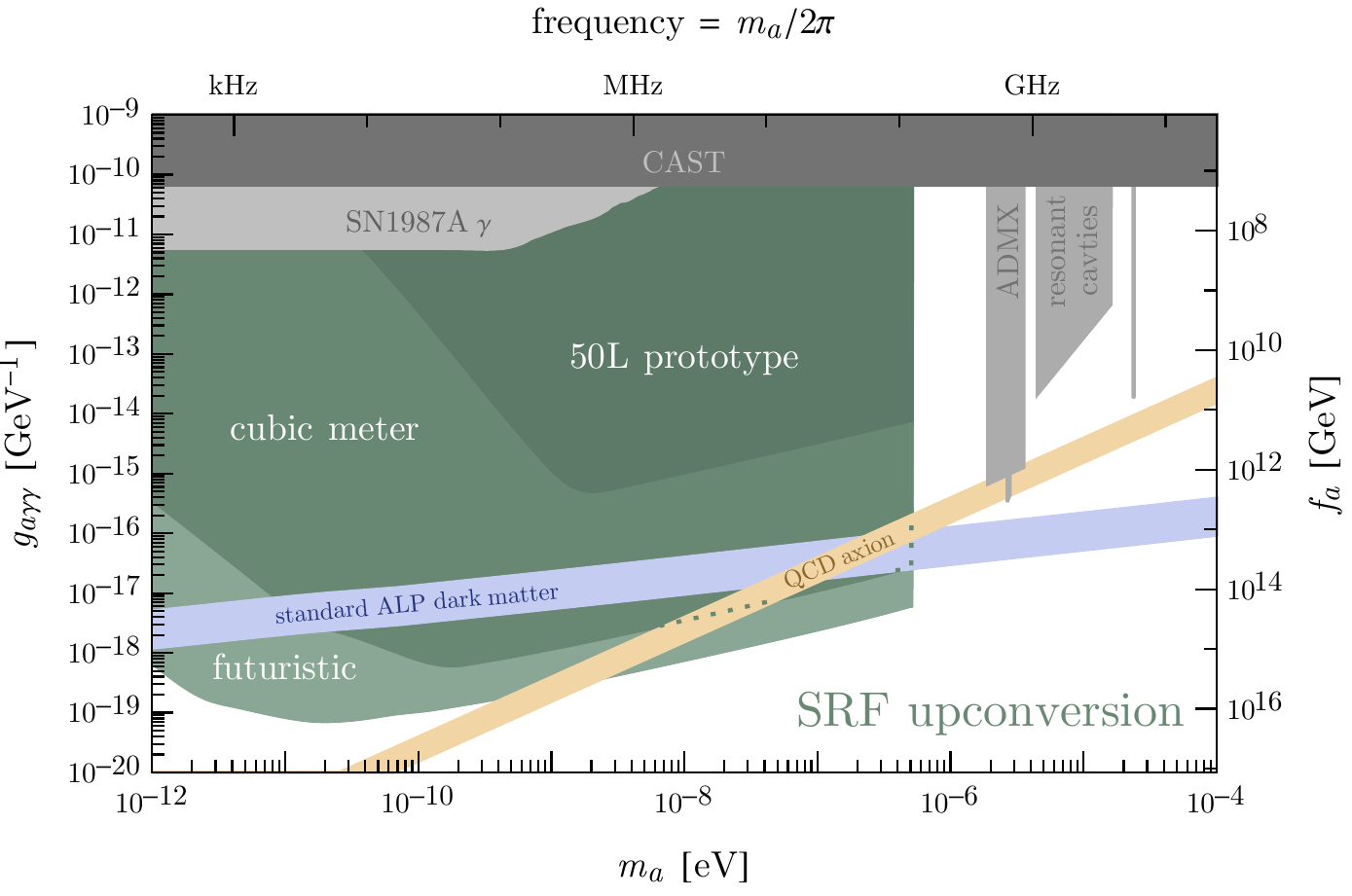}
   \end{tabular}
   \end{center}
   \caption[example] 
   { \label{fig:heterodynereach} 
 Projected  sensitivity  of  an  up-conversion axion dark matter experiment. As three representative examples, we show the projected sensitivity of a 50L prototype, a cubic meter setup, and a futuristic experiment with a total instrumented volume of $5 \ \text{m}^3$. Regions motivated by the QCD axion and ALP dark matter generated by the misalignment mechanism are shown as orange and blue bands, respectively. 
}
   \end{figure*} 

As highlighted in Refs.~\cite{Berlin:2019ahk,Lasenby:2019prg,Berlin:2020vrk}, a heterodyne search employing the oscillating electromagnetic fields of a resonant cavity completely undoes this unneeded suppression. This can be understood from the form of the axion-induced effective current introduced above in Eq.~(\ref{eq:Ja}), which in the long-wavelength limit is $\J_a \propto \B \, \partial_t a$. For a background magnetic field oscillating at a frequency of $\omega_0$, the resulting axion-induced emf is controlled by the beats of $\omega_0$ and $m_a$, i.e., $\mathcal{E}_a \propto \partial_t \J_a \propto \omega_0 \pm m_a$. Thus, compared to an experiment employing static magnetic fields, oscillating the field at $\omega_0 \sim \text{GHz}$ enhances the signal by $\text{GHz} / m_a \gg 1$ in the low-mass limit.

\begin{figure}[t]
    \centering
    \includegraphics[width=\linewidth]{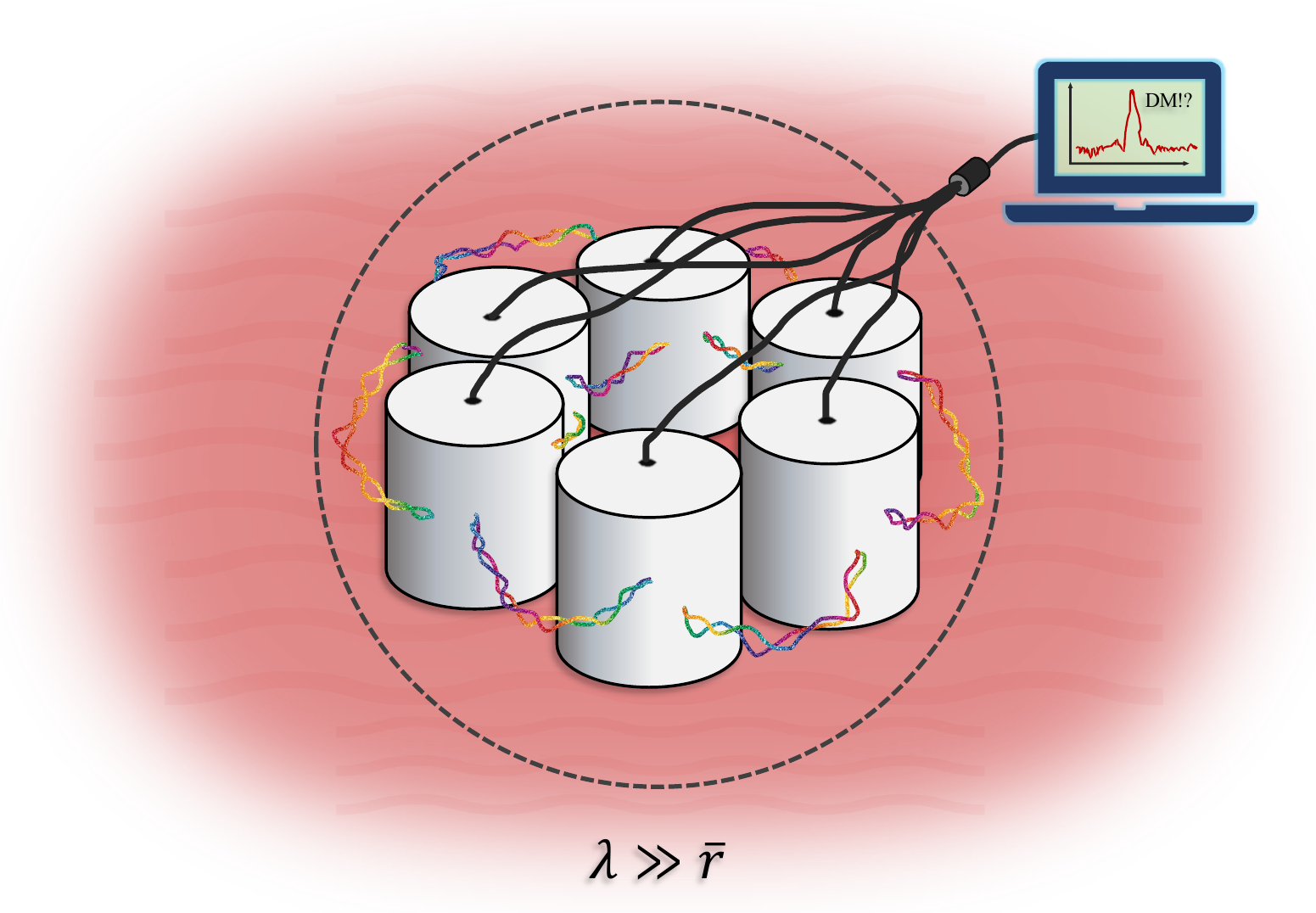}
    \caption{Illustration of entangled sensor-cavities within a network-volume of radius $\bar{r}$, taken to be much smaller than the axion coherence length, $\lambda$. Thanks to this hierarchy of scales, the entire network experiences a coherent signal that can be combined at the amplitude level. The cylinders represent the cavities, and the colored lines represent their entanglement. The solid black lines represent joint processing of the measurement to search for dark matter.}
    \label{fig:cavity_network_pic}
\end{figure}

This enhancement carries over directly to the reach of a proposed heterodyne search utilizing a single SRF cavity~\cite{Berlin:2019ahk,Lasenby:2019prg,Berlin:2020vrk}. A schematic of the setup is shown in Fig.~\ref{fig:heterodynesetup}. A cavity is pumped at a frequency $\omega_0 \sim \text{GHz}$. The axion field interacts with the pump mode magnetic field, resonantly driving power into a signal mode of nearly degenerate frequency $\omega_1 \simeq \omega_0 \pm m_a$ and distinct spatial geometry, provided that the pump $B$-field and signal $E$-field have a large overlap $\B_0 \cdot \E_1$ integrated over the volume of the cavity (which is indeed the case when the pump and signal modes are, e.g., low-lying TE and TM modes). A scan over possible axion masses can be achieved by slightly perturbing the geometry of the cavity, thus modifying the frequency splitting $|\omega_1 - \omega_0|$ to match a wide range of axion masses. Compared to a static-field LC circuit experiment of comparable size, the SNR of a thermal-noise limited SRF heterodyne setup is enhanced by
\begin{equation}
\frac{\text{SNR}_\text{SRF}}{\text{SNR}_\text{LC}} \sim \frac{\omega_1}{m_a} \, \bigg(\frac{Q_\text{SRF}}{Q_\text{LC}}\bigg)^{\frac{1}{2}} \bigg(\frac{B_\text{SRF}}{B_\text{LC}}\bigg)^{2} \bigg(\frac{T_\text{LC}}{T_\text{SRF}}\bigg)^{\frac{1}{2}} 
,
\end{equation}
where $Q$, $B$, and $T$ are the quality factor, magnetic field strength, and temperature of the corresponding setup, respectively. Taking representative experimental parameters in the above expression, the larger quality factor of SRF cavities approximately compensates for the smaller magnetic fields and larger temperatures achievable in a superconducting heterodyne setup, leaving a residual enhancement $\text{SNR}_\text{SRF} / \text{SNR}_\text{LC} \sim \mu \text{eV} / m_a \gg 1$ that becomes increasingly important at smaller axion masses. Furthermore, since the frequency of the axion signal $\omega_1 \sim \omega_0 + m_a$ is approximatley independent of the axion mass in the long-wavelength limit, GHz cavities can efficiently couple to extremely slowly oscillating axion fields, opening up sensitivity to axion masses below a kHz, a region typically inaccessible with static field resonators.

\begin{figure}[t]
    \centering
    \includegraphics[width=\linewidth]{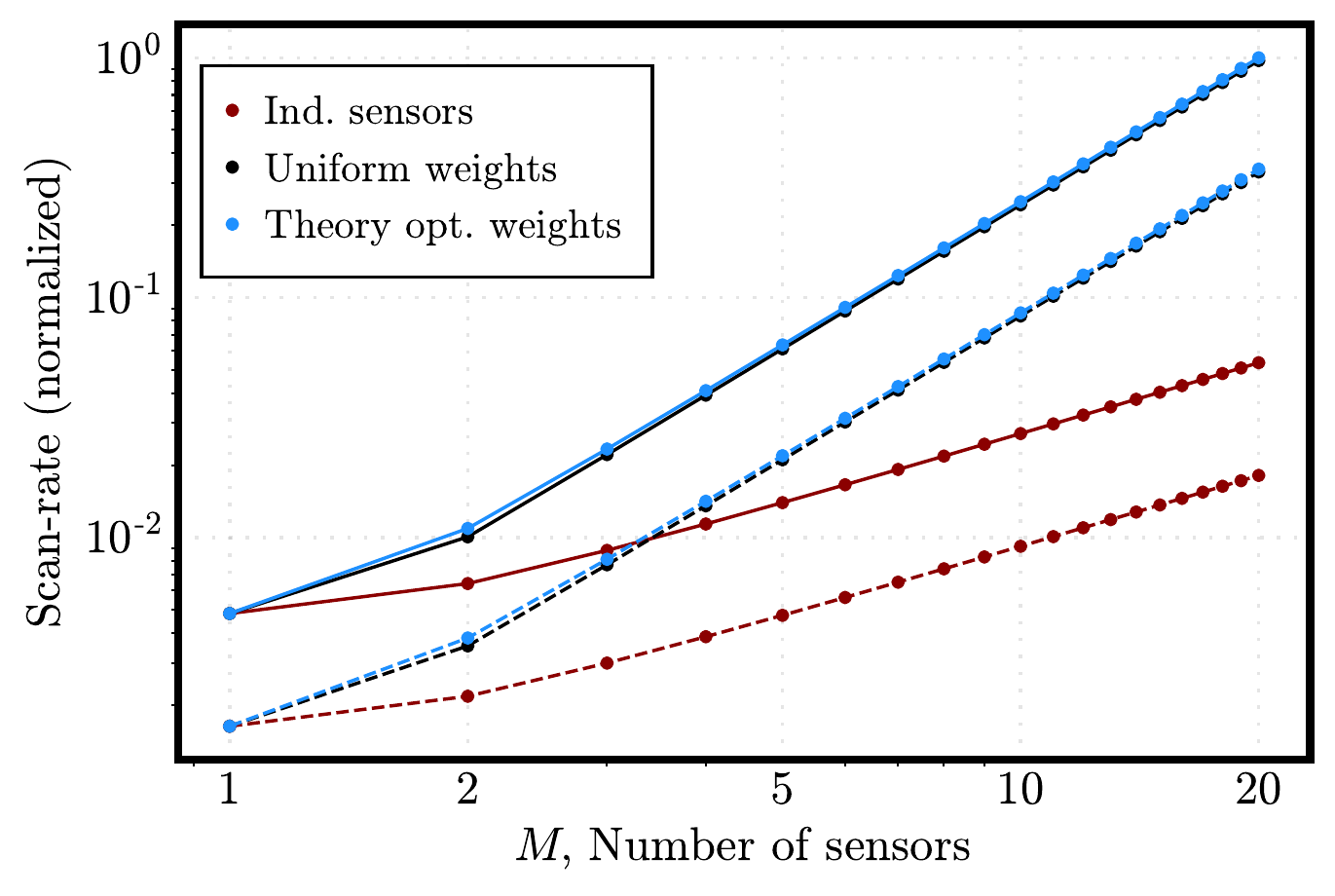}
    \caption{Scaling of the (normalized) scan-rate with the number of sensors $M$ in the network (log-log plot). Cavities in the network differ by their intrinsic linewidths, which fall within the interval $\gamma_{\ell}\in[1,3]$ (normalized by smallest linewidth). Solid lines correspond to squeezed input at gain $G=4$. Dashed lines correspond to the quantum limited setup (zero squeezing). For a network of independent sensors, each sensor must be independently squeezed. Observe quadratic scaling of the scan rate with $M$ for a coherent network of sensors versus linear scaling with $M$ for a network of independent sensors (i.e., the slope of the former is about twice that of the latter).}
    \label{fig:scanrate_dqs}
\end{figure}

The projected sensitivity to the axion parameter space is shown in Fig.~\ref{fig:heterodynereach} for three choices of experimental parameters, including noise estimates that are anchored to previous experimental measurements, such as those performed for the MAGO gravitational wave experiment~\cite{Bernard:2001kp,Bernard:2002ci,Ballantini:2005am}. A wider range of experimental parameters is shown in Refs.~\cite{Berlin:2019ahk,Berlin:2020vrk}. The projection for the 50L prototype assumes a peak pump-mode magnetic field of $B_\text{pump} = 0.2 \ \text{T}$, an intrinsic quality factor of $Q = 10^{10}$, and no attenuation of mechanical vibrations. The sensitivity of the cubic meter setup assumes the same magnetic field, a larger quality factor of $Q = 10^{12}$, and 100 dB of vibrational attentuation. Finally, the projections for the futuristic setup assumes a total integrated volume of $5 \ \text{m}^3$, $B_\text{pump} = 0.4 \ \text{T}$, $Q = 10^{12}$, and 130 dB of vibrational attenuation. 

These estimates indicate that noise induced by mechanical vibrations are likely to limit the sensitivity for $m_a \lesssim \text{kHz}$, while a thermal-noise limited setup is potentiallly attainable for $m_a \sim \text{kHz} - \text{GHz}$. In particular, a thermal-noise limited cubic-meter sized cavity woud be sensitive to the QCD axion for $m_a \sim 100 \ \text{kHz} - 100 \ \text{MHz}$. As mechanical noise increases rapidly at lower frequency splittings, there is strong motivation to consider active or passive attenuation of vibrations stemming from the immediate environment, such as the cooling apparatus, as was considered in Ref.~\cite{Ballantini:2005am}. Regardless, a wide range of uncharted axion DM parameter space remains attainable without any active attenuation and for modest experimental assumptions.

\subsection{A Quantum Network of Haloscopes}

The operation of haloscopes in ultra-coherent cavities and quantum techniques presents new opportunities to enhance the scan rate. One such example has been shown by the HAYSTAC experiment~\cite{Backes2021}. There, the SNR and the scan rate for axion DM have been enhanced with a quantum metrology protocol~\cite{Zheng:2016qjv}. In particular, a squeezed vacuum state injected into an RF cavity and homodyne detection along the squeeze direction together allow for an improved SNR and a faster scan.

Recently, it was shown that the performance of a quantum network could be utilized further to improve axion DM searches~\cite{Brady}. In particular, a network of microwave cavities that is local will experience a coherent axion signal if the size of the network is less than the axion coherence length, of order $10^3 \, m_a^{-1}$, as illustrated in Fig.~\ref{fig:cavity_network_pic}. However, the noise in the network will be incoherent among the network nodes. One 
can make use of distributed squeezed states~\cite{zhuang2018} to exploit the coherent nature of the DM signal. The results of~\cite{Brady} indicate that combining quantum resources (squeezing) in a distributed-network setting can allow for a scan that is faster by a factor of the square number of network nodes in the ideal case. The improvement is enabled by adding the signal at the amplitude level rather than adding powers in the classical network case.
The improved scan rate is shown in Fig.~\ref{fig:scanrate_dqs} as compared with a classical network, both with squeezing and without squeezing (i.e., at the quantum limit).

\section{High Frequency Gravitational waves}

\begin{figure}[t]
   \centering
   \includegraphics[width=0.8\columnwidth]{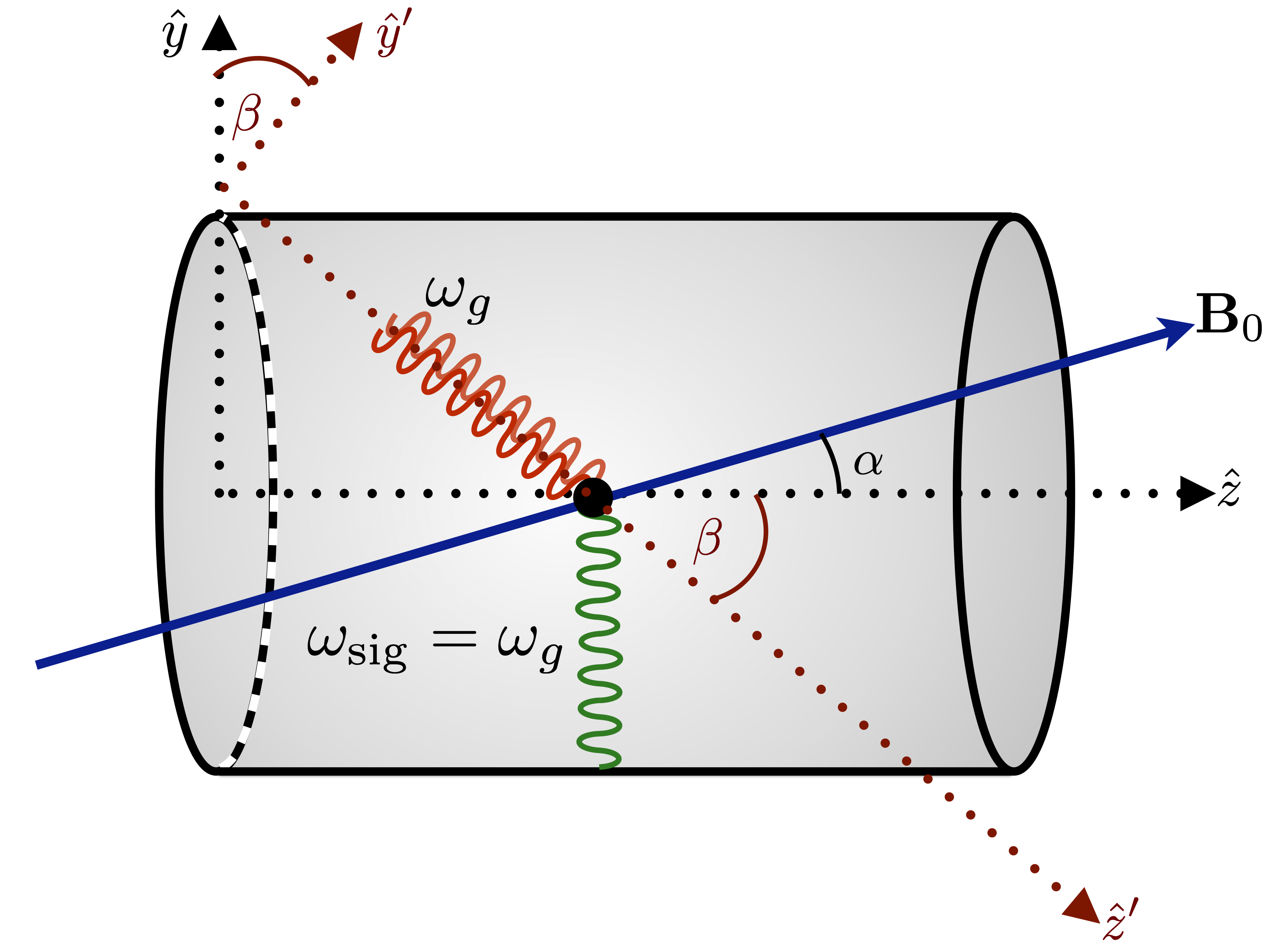}
   \caption{A gravitational wave (red) interacts with a static external $B$-field (blue), resonantly exciting a cavity mode (green).}
   \label{fig:GW_cavity_alpha_beta} 
\end{figure}

Since gravity couples to all forms of energy, there is an interaction between gravitational waves (GWs) and photons~\cite{gertsenshtein1962wave,Zeldovich:1972mn,zel1973electromagnetic}. A direct consequence of this coupling is that microwave cavities can also be used to detect high-frequency GWs. As with the static-field axion searches described above, an incoming GW can be thought of as sourcing an effective current through its interaction with a background magnetic field $B_0$. This effective current then sources a small signal field $B_{\rm sig} \sim h \, B_0$, where $h$ is the characteristic strain of the GW. Such signals were recently investigated in Ref.~\cite{Berlin:2021txa}. 

GWs are typically classified into two different categories: stochastic GW backgrounds, and GWs with a preferred direction of propagation and well-defined frequency.\footnote{Such sources are often transient, and thus the frequency typically evolves in time, with the particular timescale depending on the source of emission.} We concentrate on the latter case, since the detection prospects are typically better. A detailed theoretical analysis of the sensitivity to stochastic signals is ongoing. 

\begin{figure}[t]
   \center
   \includegraphics[width=\columnwidth]{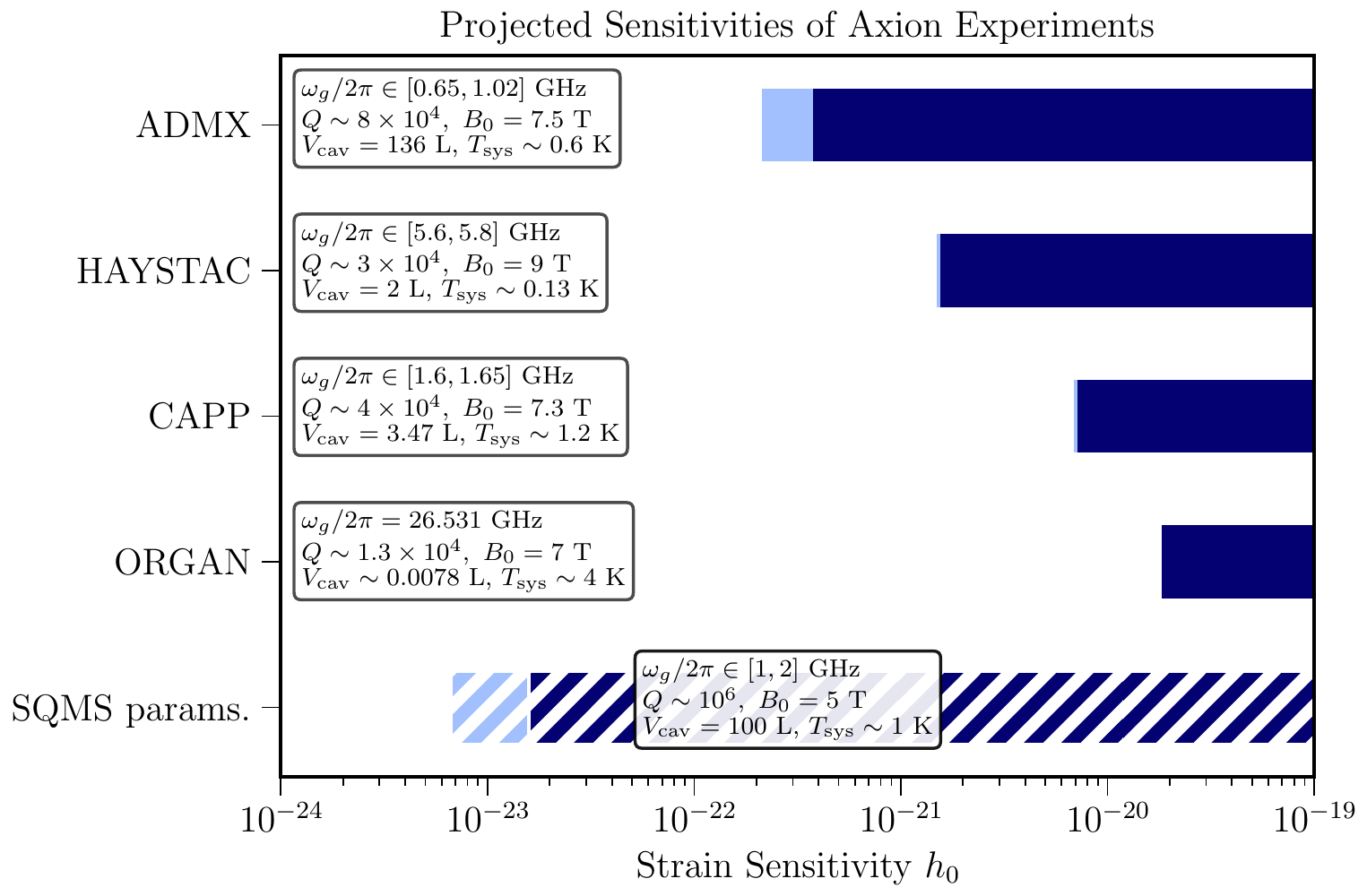}
   \caption{The projected sensitivity of axion cavity haloscope experiments to high-frequency coherent gravitational waves.}
   \label{fig:GW_sensitivity} 
\end{figure} 

The setup we focus on in this section is a cavity that is permeated by a static external $B$-field. The incoming GWs can interact with the static $B$-field of the cavity and excite a specific cavity mode, as shown in Fig.~\ref{fig:GW_cavity_alpha_beta}. Such a setup is most sensitive near resonance, so that the resonant frequency matches the GW frequency, $\omega_{\rm sig} \simeq \omega_g \sim \text{GHz}$.
Possible sources of GWs in this frequency range are binary mergers of ultralight compact objects with mass $M \lesssim 10^{-11}M_\odot$ and GWs emitted by the superradiant axion cloud around spinning primordial black holes, corresponding to masses of $M\sim 10^{-4}\, M_\odot$. For a review on high frequency GW sources and other recent proposals, see, e.g. Ref.~\cite{Aggarwal:2020olq,Domcke:2022rgu}.

Following Ref.~\cite{Berlin:2021txa}, the sensitivity of the described setup to the GW strain $h$ is
\begin{eqnarray}
h &\gtrsim& 3 \times 10^{-22} \times \bigg( \frac{1 \ \text{GHz}}{\omega_g / 2 \pi} \bigg)^{\frac{3}{2}} \bigg( \frac{0.1}{\eta_n} \bigg) \bigg( \frac{8 \ \text{T}}{B_0} \bigg) \bigg( \frac{0.1 \ \text{m}^3}{V} \bigg)^{\frac{5}{6}}\nonumber\\ &~&\bigg( \frac{10^5}{Q} \bigg)^{\frac{1}{2}} \bigg( \frac{T_\text{sys}}{1 \ \text{K}} \bigg)^{\frac{1}{2}} \bigg( \frac{\Delta \nu}{10 \ \text{kHz}} \bigg)^{\frac{1}{4}} \bigg( \frac{1 \ \text{min}}{t_\text{int}} \bigg)^{\frac{1}{4}}
~,
\label{eq:sensitivity_estimate}
\end{eqnarray}
where $\omega_g$ is the GW frequency, $\eta_n$ the coupling coefficient defined below, $B_0$ the magnitude of the external $B$-field, $V$ the cavity volume, $Q$ the cavity quality factor, $T_\text{sys}$ the system temperature, $\Delta \nu$ the signal bandwidth, and $t_\text{int}$ the integration time. In Fig.~\ref{fig:GW_sensitivity} we show the sensitivity of cavity experiments originally designed to detect axion DM. All experiments shown (except that labelled ``SQMS" which corresponds to a cavity under development) only need to reanalyze data already recorded for axion DM searches in order to set limits on GWs in the GHz regime.

The coupling coefficient $\eta_n$ in the sensitivity estimate of Eq.~\eqref{eq:sensitivity_estimate} is defined as
\begin{eqnarray}
\label{eq:formfactor}
\eta_n \equiv \frac{\Big| \int_{V} d^3 \mathbf{x} ~ \bm{E}_n^* \cdot \hat{\boldsymbol{j} }_{ +, \times} \Big|}{V^{1/2} \, \big( \int_{V}   d^3 \mathbf{x}~|\bm{E}_n|^2 \big)^{1/2}}
~,
\end{eqnarray}
where the index $n$ stands for a generic mode of the cavity, $\bm{E}_n(\bm{x})$ the spatial cavity mode function, and $\hat{\boldsymbol{j} }_{ +, \times}$ the effective current induced by the GW. In particular, $\hat{\boldsymbol{j} }_{ +, \times}$ depends on the GWs polarization, which is indicated with a $+$ or $\times$ sign. The full definition can be found in Ref.~\cite{Berlin:2021txa}.
The coupling coefficient is best evaluated in the local inertial frame of the cavity.
This frame
is often referred to as the proper detector frame or Fermi normal coordinate system~\cite{Fermi:1922abc,Fermi:1922def,Manasse:1963zz,Marzlin:1994ia,Rakhmanov:2014noa}. Furthermore, resonant cavities operate in the regime where the GW frequency is comparable to the inverse geometric size of the cavity, i.e., $\omega_g L_{\rm det} \sim 1$. In this case, evaluating the signal strength requires the resummation of an infinite series of derivatives acting on the Riemann tensor~\cite{Berlin:2021txa}. 

\begin{figure} [t]
    \centering
     \includegraphics[width=0.8\columnwidth]{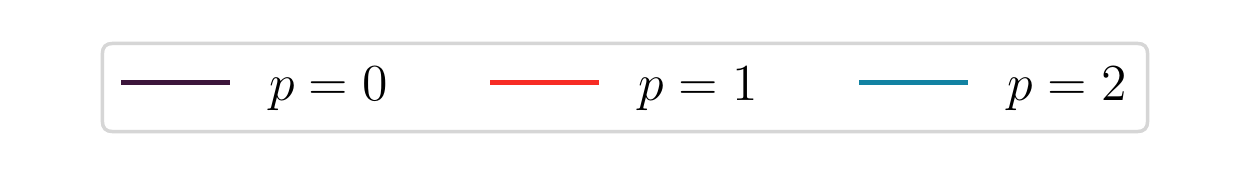}
    \includegraphics[width=\columnwidth]{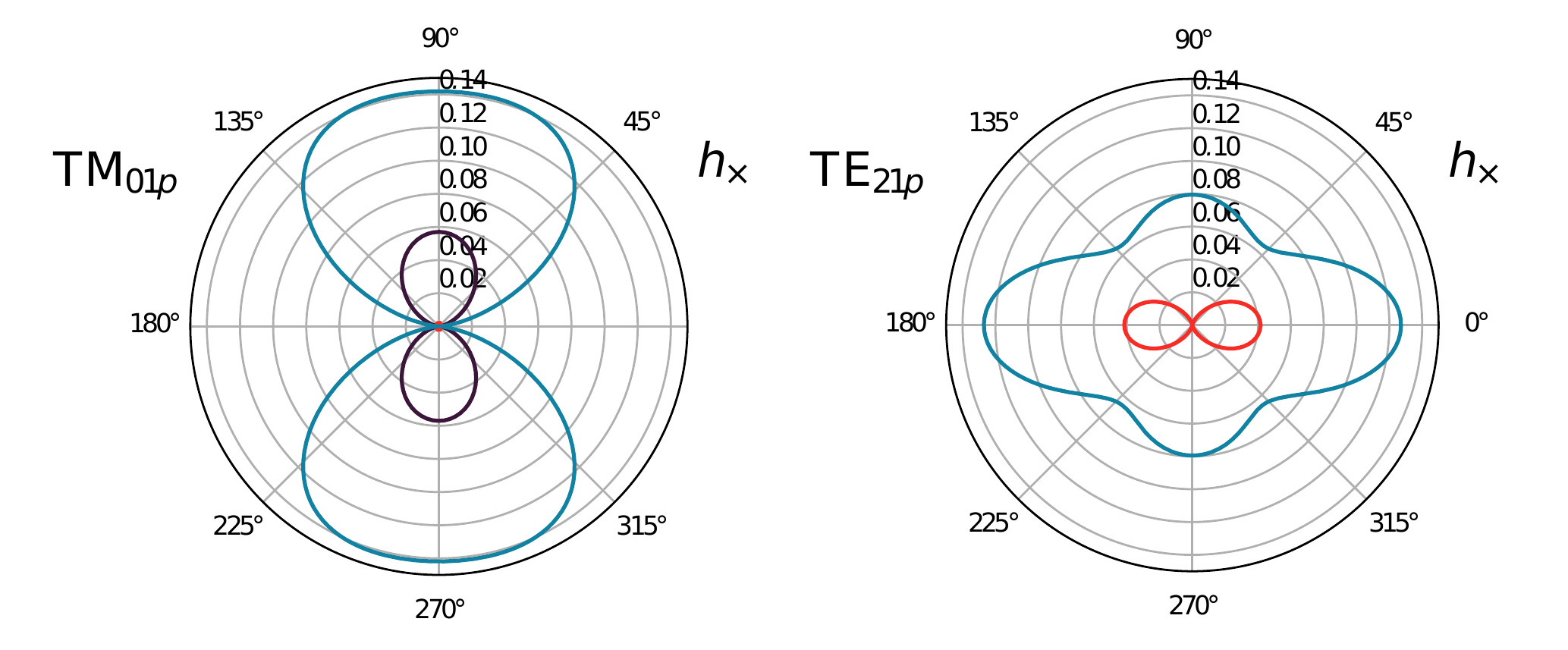}
   \caption{Polar plot for the coupling coefficient $\eta_n$ for TM$_{01p}$ and TE$_{21p}$ modes. The polar angle is $\beta$, the orientation of the GW relative to the bore of the cavity, as depicted in Fig.~\ref{fig:GW_cavity_alpha_beta}.}
   \label{fig:GW_coupling_coefficient} 
\end{figure}

In Fig.~\ref{fig:GW_coupling_coefficient} we show the coupling coefficient as a function of the incidence angle $\beta$ of the GW. The external $B$-field is taken to be coaxial with the cavity bore ($\alpha=0$). We consider two modes of a cylindrical cavity: TM$_{01p}$ (left) and TE$_{21p}$ (right). The coupling coefficient is non-zero in the TM$_{010}$ mode used by most axion haloscopes, such that already-recorded data can be reanalysed to search for GWs. 
From Fig.~\ref{fig:GW_coupling_coefficient} it is clear that 
the detailed spatial structure and mode-dependence of the signal enables sensitivity to the direction and polarization of the GW.

High-frequency GWs can also be detected by the interaction of GWs with a pumped cavity mode, analogous to the upconversion axion search in Sec.~\ref{sec:UpConversion}. In this case, the GWs induce the transition of photons from the pumped cavity mode into an empty signal mode~\cite{Braginskii:1973vm,Grishchuk:1975tg,Pegoraro:1977uv,Pegoraro:1978gv,Caves:1979kq,Bernard:2001kp,Bernard:2002ci,Ballantini:2003nt,Ballantini:2005am,Mensky:2009zz}. A detailed analysis taking into account the proper detector frame, and interactions of the GW with mechanical modes of the cavity is currently ongoing. Since this setup shares many features with multi-mode axion DM searches, such as in Sec.~\ref{sec:UpConversion}, the R\&D for these two goals is synergistic.

\section{Capabilities and Infrastructure}

SRF cavities, as we show above, provide a rich new arena to explore new physics. In this section we give examples of new capabilities and new supporting infrastructure. In particular we present ongoing and future research on optical-RF transduction, which would have an impact on single photon sensing and senor networks. We also comment on opportunities to scale-up cavity based sensing experiments, including a planned cryogenic platform at SQMS in Fermilab which would be the world's biggest of its kind.

\subsection{Enhanced sensitivity by High-Efficient Microwave-Optical transduction}

The strong effort to utilize SRF cavities for QIS at SQMS has inspired a research on high efficiency microwave-optical transduction based on high-$Q$ SRF resonators. A high  RF-to-optical transduction efficiency may enhance new physics searches.
For example, up/down photon conversion at high efficiency can extend new physics particle searches over a wider energy spectrum, with the possibility to detect single photons in both microwave and optical regimes, as well as performing measurements below the SQL. Up/down photon conversion may also enable highly sensitive axion and dark photon haloscope searches in the THz regime. One possibility is to let the axion or dark photon convert to a SM photon in a THz cavity. Transduction would then allow this signal to accumulate coherently in a much larger GHz range microwave cavity, greatly enhancing the SNR of a possible DM signal.

Fermilab is supporting a preliminary study on quantum transduction through the Laboratory Directed R\&D (LDRD) program. We are exploring hybrid coherent resonance systems and bi-directional quantum transduction schemes to up/down- convert the microwave information to/from the optical regime and enhance the conversion efficiency at the quamtum threshold.




Non-centrosymmetric crystals are used to create interactions between microwave and optical fields using photonic RF three-wave-mixing processes or electro-optic modulation. An electric field, applied on these crystals, modulates the refractive index and the incident optical field linearly with the RF voltage ($\chi^{(2)}$), which is known as the Pockels effect. In microwave-optical transduction, the self-heterodyne techniques have the advantage of eliminating the local oscillator (LO) used in conventional super-heterodyne mixers, reducing therefore complexity, power consumption, added noise, and thermal quasiparticle poisoning. 
In nonlinear electro-optic materials, the RF (i.e., the excitation mode of the resonant cavity) and the sidebands are mixed through the second order nonlinearity in the transmitted optical power: $P_{\text{opt}}\propto V^2_{\text{RF}}$. We are interested in detecting the sidebands of the optical signal. The fraction of transmitted optical power related to the sideband of the RF signal is proportional to the optical field $|E_{\text{opt}}|^2$. If the sideband of the microwave signal $V_{\text{RF}}$ is modulated in phase $\varphi_b(t)$, this phase modulation is also detected as $P_{\text{opt}}(t)\propto(1+\cos(\varphi_b(t))+\cos(2\varphi_b(t)))$. 






The integration between different technologies in a Hybrid Quantum System (HQS) is a major challenge which goes beyond the design, implementation, and the control of single components. The Hamiltonian to model this system electro-optic interaction in a triple-resonance scheme can be reduced to the following form:
\begin{equation}
    \hat{H}= \hslash g_{eo} (pa^\dag + p^\dag a) (b+b^\dag),
\end{equation}
the pump optical mode ($p$) is driven to coherently couple the optical signal mode ($a$) with the microwave mode ($b$), with the electro-optic (EO) coupling strength $g_{eo}$~\cite{kurizki2015quantum}. 


\begin{figure}[!t]
  \begin{subfigure}{0.4\linewidth}
    \centering
    \includegraphics[height=3 cm]{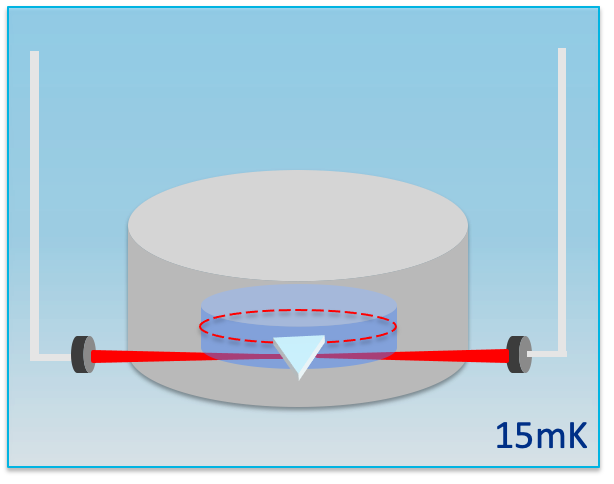}
    \caption{}
    \label{fig:transduction_cavity_block_diagram}
    \end{subfigure}
     \begin{subfigure}{0.4\linewidth}
    \centering
    \includegraphics[height=3 cm]{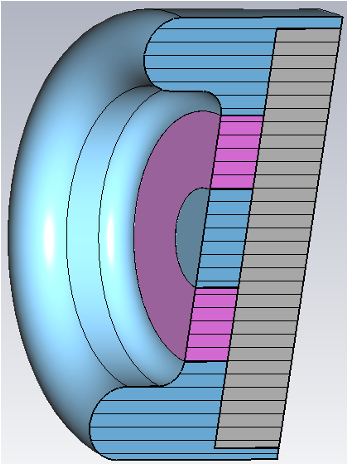}
    \caption{}
    \end{subfigure}
    \caption{(a) Bulk Niobium cavity, superconducting at cryogenic temperature coupled to $\text L \text i \text N \text b \text O_\text 3$ whispering-gallery mode crystal, with a prism for the optical coupling with fiber-optics, operating in a dilution refrigerator at 15 mK. (b) Cross-section of a hybrid microwave-optical cavity.}\label{fig:transduction_cavity}
\end{figure}

Limiting factors in the attainment of high-efficiency in the microwave-optical transduction  are:
\begin{itemize}
    \item Quality factor of the microwave cavity -- While the optical $Q_a$ can reach $10^7$, microwave $Q_b$ is limited in the current schemes.
    \item Single-photon microwave-optic coupling coefficient ($g_{eo}$) -- This paramter is determined by the overlap between microwave and optical fields. 

    \item Pump power -- High pump power leads to overheating of the device, and thermal quasiparticles poisoning.
\end{itemize}

\begin{figure}[ht]
   \centering
  \begin{subfigure}{0.75\linewidth}
 \includegraphics[height=4.5 cm]{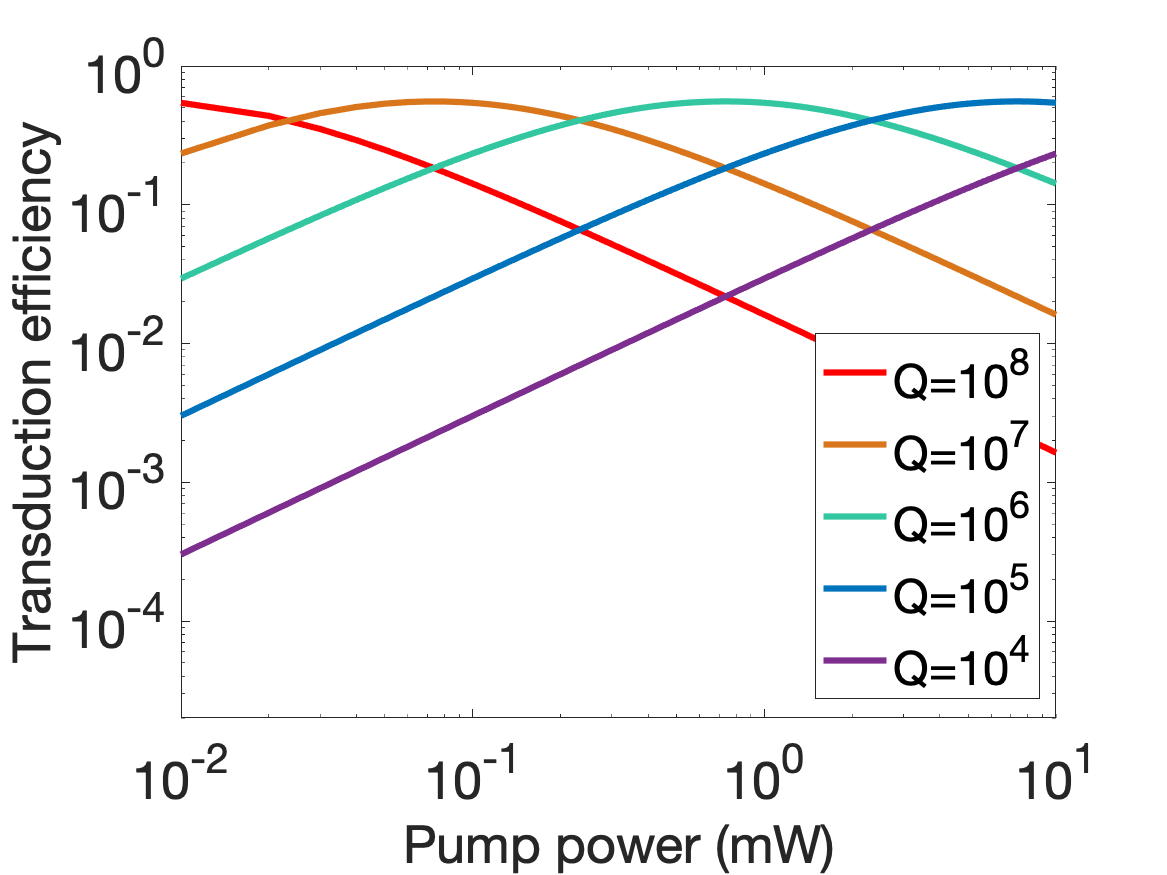}
    \caption{}
   \label{fig:transduction_efficiency}
  \end{subfigure}
      \begin{subfigure}{0.75\linewidth}
   \includegraphics[height=4.5 cm]{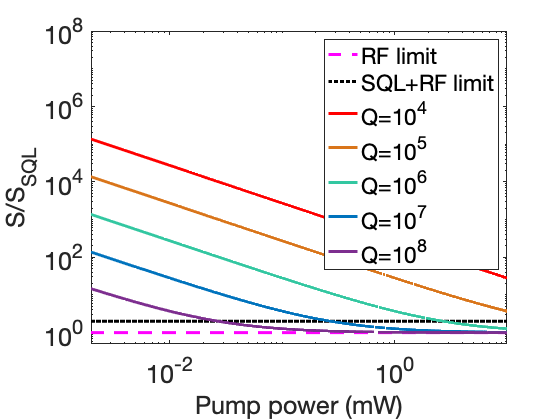}
  \caption{}
 \label{fig:sql}
 \end{subfigure}
    \caption{(a) Quantum transduction efficiency ($\eta_i$) and (b) Spectral density (S) with reference to the SQL, with back-action noise cancellation. Both plots evaluate the parameters with increasing RF quality factors (Q).} \label{fig:transduction_parameters}
\end{figure}

The bi-directional efficiency depends on the  microwave-optical cooperativty and losses:
\begin{equation}\label{eq_efficiency}
    \eta=\frac{\kappa_{a,ex}}{\kappa_a}\frac{\kappa_{b,ex}}{\kappa_b}\times\frac{4C}{(1+C)^2}
    ~~,~~
    C=\frac{4n_p{g}_{eo}^2}{\kappa_a\kappa_b}
\end{equation}
where, for a generic $m$ mode ($m=a,b$), $\kappa_{m}$ is the total loss rate, consisting of both the intrinsic and external losses: $\kappa_m=\kappa_{m,ex}+\kappa_{m,i}$. $C$ is the cooperativity between the optical and the microwave modes, which depends on the photon number in the pump mode ($n_p$). The second term of Eq.~\ref{eq_efficiency} is the internal efficiency: $\eta_i= 4C / (1+C)^2$.

We evaluate the single photon coupling as 
\begin{equation}
    g_{eo}=\frac{1}{8} n^2 \omega_p r_{33} E_\text{max} \sqrt{\frac{\hbar \omega_m}{W}}
\end{equation}
where $n$ is the reflective index of the non-centrosymmetric crystal, $\omega_p$ is the optical pump power, $r_{33}$ is the electro-optic coefficient, $E_\text{max}$ is the maximum electric field that overlaps with the optical mode at the edge of the crystal, $\omega_m$ is the microwave frequency, and $W$ is the stored energy in the microwave cavity. 
Lithium niobate ($\text L \text i \text N \text b \text O_\text 3$) is among the best materials exhibiting the Pockels effect, with
a large electro-optic coefficient ($r_{33} = 31 \ \text{pm}/ \text{V}$ at 9 GHz) and low optical losses. In Fig.~\ref{fig:transduction_cavity}, we show the diagram of a microwave-optical device based on bulk resonators, along with the 3D model of a possible SRF hybrid cavity. The design focuses on the optimal overlap between the microwave and the optical fields, maximizing the quality factor and $E_\text{max}$, to achieve high efficiency and cooperativity, with low added noise. Analyzing microwave simulations we observe that hybrid devices with higher $Q$ achieve maximum conversion efficiency at lower pump powers, as shown in Fig.~\ref{fig:transduction_efficiency}. Therefore longer coherence in the microwave cavity will ensure low-noise for few-photon quantum transduction, which is different from the conventional trade-off between efficiency and noise~\cite{rueda2016efficient,hease2020bidirectional,holzgrafe2020cavity}.

A combination of high-efficiency and quantum noise reduction schemes can be proposed to increase the sensitivity, such as squeezed light and back-action noise cancellation techniques~\cite{nazmiev2021back,tsang2010cavity}. Fig.~\ref{fig:sql} shows spectral densities with reference to the SQL. A back-action noise evading scheme is applied to measure a combination of in-phase and quadrature RF signals, achieving a sensitivity better than the SQL. We observe that by increasing the $Q$ in the microwave cavity, the SQL is reached at lower pump powers.



\subsection{Cryogenic Platform for Scaled-up Sensing Experiments}

The SQMS center at Fermilab is developing a cryogenic platform capable of reaching millikelvin temperatures in an experimental volume of 2~meters diameter by 1.5~meters in height~\cite{Hollister:2021lhg}. 
The platform is designed to host a three-dimensional qubit architecture based on SRF technology, as well as sensing experiments. 

\begin{figure}[t]
    \centering
    \includegraphics[width=9cm]{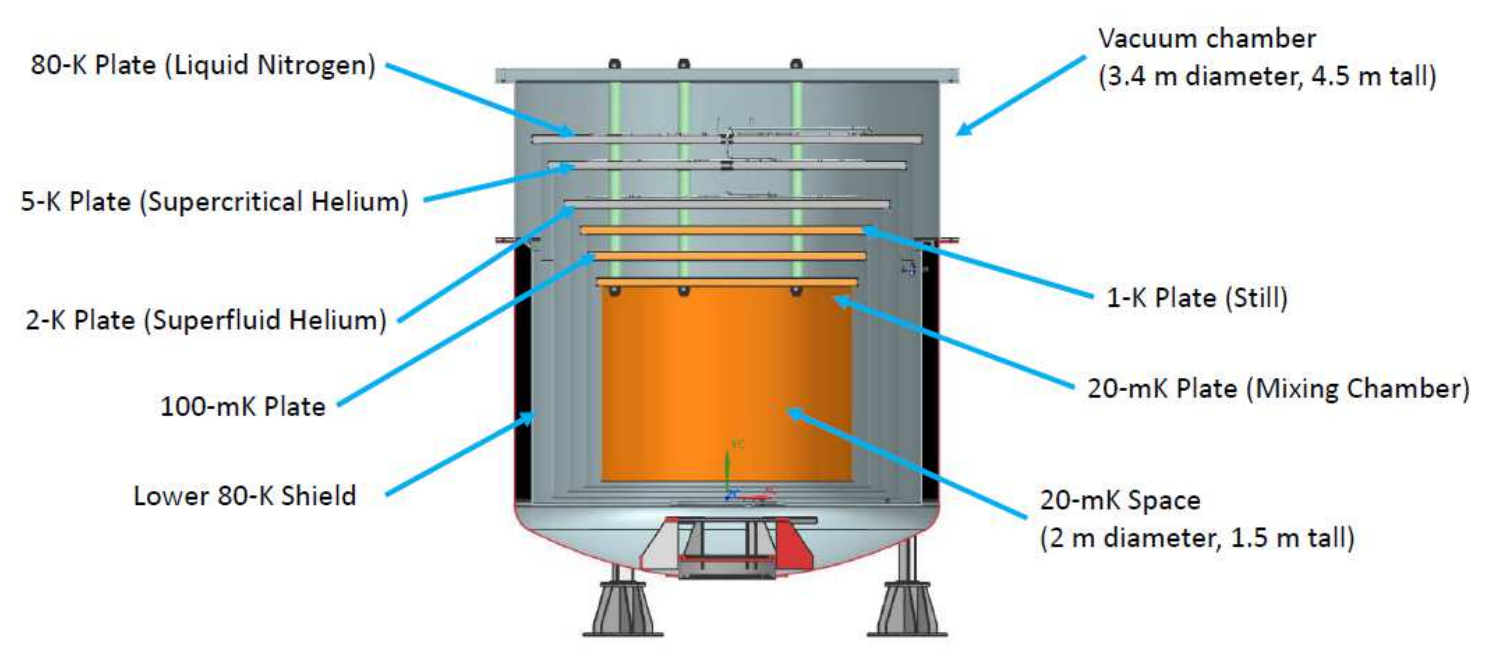}
    \caption{A CAD rendering of the planned SQMS cryogenic platform for quantum computing and sensing. Specifications of the system are presented in~\cite{Hollister:2021lhg}.}
    \label{fig:colossal}
\end{figure}

The system, which is sketched in Figure~\ref{fig:colossal}, will consist of several cooling stages of which we quote here specifications for two. The 20~mK space will consist of a cooled volume of 4.7~m$^3$ and with a cooling power of 300~$\mu$W, while the 2K stage (superfluid Helium) will simultaneously have a cooling power of order 30~Watt. The system can be operated as a 2-K platform only, in which configuration the cold volume will be of order 10~m$^3$. 
The specifications for other stages of the platform are listed in~\cite{Hollister:2021lhg}.     

\section{Outlook and future R\&D}

The development of high quality superconducting cavities for accelerators presents some low hanging fruit to explore new fundamental physics. Off-the-shelf systems and existing test facilities are already probing new ground, for example, with Dark SRF's dark photon search.

Taking full advantage of high quality systems and reaping the full set of physics opportunities presented in this document requires dedicated R\&D. An incomplete list of research directions that are planned or needed include: 
\begin{itemize}
\item Material science for improved $Q$ -- The new physics searches outlined here benefit from high quality factors. Ongoing research aimed at enhancing the coherence time of SRF cavities for use as quantum devices (e.g., Ref.~\cite{Romanenko:2018nut}) is synergistic with the sensing goals (including achieving high $Q$ when a cavity is in a multi-tesla magnetic field).

\item Quantum engineering of cavity states -- Sensing goals can be  aided by preparing nontrivial  quantum states in electromagnetic cavities (e.g., squeezed states~\cite{Zheng:2016qjv,Malnou:2018dxn,HAYSTAC:2018rwy}). As we discussed, further enhancement can be achieved with a network of entangled cavities~\cite{Brady}.

\item Cavities and Qubits -- The integration of a qubit into an ultra high cavity may enable new schemes for quantum computing~\cite{SQMS-whitepaper} and synergetically allow for employing a photon counting non-demolition measurement for DM searches, as exemplified in Ref.~\cite{Dixit:2020ymh}. For certain DM search schemes, it would also be beneficial to have qubits that can operate successfully even in high magnetic fields.

\item Nonlinearities in superconducting cavities -- Searches with multiple cavity modes, for axions or gravitational waves, depend on observing a faint signal in the presence of an excited mode (or modes) in the background. Nonlinear effects within the cavity walls can mimic such a signal, particularly if the signal mode is near a harmonic. Nonlinear effects have recently been studied and shown to depend on both the disorder parameter and the temperature of the superconductor~\cite{Sauls:2022sax}. Experimental studies to measure and mitigate such nonlinearities are well-motivated.
\item Nonlinearites due to TLS and other impurities -- Nonlinear effects can also come from impurites on the cavity walls. Two-level systems (TLS) are a known source of loss in superconducting devices and are likely to induce nonlinear current responses at some level. Measurements and estimates of TLS-induced nonlinearity is required. Mitigating TLS altogether, or minimizing their overlap with cavity modes is well-motivated for our sensing goals, as well as to enhance the coherence of superconducting devices for quantum applications.
\item Frequency stability -- the cavity frequency can change during measurements, for example due to microphonics (i.e., mechanical vibrations, helium pressure fluctuations, etc.) or Lorentz force detuning. Developing methods for mitigating the effects of these phenomena will be especially important when operating with a narrow bandwidth, which will help to increase SNR.

\item Cavity and antenna design -- cavity search schemes often involve different modes, and in many cases, it is beneficial to use different antennas to couple to different modes, so that each couples strongly to the mode of interest, but weakly to the others. This can help avoid nonlinear excitation and other phenomena that can cause a spurious signal.

\item Frequency tuning -- to scan over a mass range of interest, tuners are used to vary the resonant frequency of cavities. There are a variety of tuning systems that are implemented, with different mechanisms, tuning ranges, tuning speeds, and other qualities such as hysteresis. In millikelvin systems, a critical parameter to consider is heat generation. A tuner that uses an electrical motor may overwhelm the cooling of such a system. Even a piezo tuner may still generate enough heat to raise the temperature of the cavity after a tuning step, requiring cooldown time. Innovation in low dissipation, wide range frequency tuning systems would be beneficial.

\item Vibration noise -- New tools for improving isolation of systems under test from sources of vibrations will be helpful for future experiments. 

\item RF sources -- RF noise sources can include not just thermal photons, but also crosstalk, amplifier noise, and nonlinearities. New techniques and tools to reduce these sources of noise will be important, in addition to diligence when it comes to established practices to limit RF noise. Many such techniques have already been developed in the context of precision metrology, and their incorporation into SRF cavity searches represents an important opportunity for collaboration between these fields.

\item Quantum transduction -- Microwave cavities with long coherence will increase the microwave-optical transduction cooperativity and efficiency, ensuring low-noise conversion at the quantum threshold. Sensitivity below the SQL is also enhanced by the higher microwave $Q$. The up/down conversion may enable highly sensitive axion and dark photon haloscope searches in the THz regime.

\end{itemize}

We conclude that advances in SRF and quantum technology present exciting opportunities to probe fundamental physics. We have presented examples, including searches for new particles, searches for DM, and new approaches to detection of gravitational waves. The first searches employing SRF cavities are underway, both in the classical and quantum regime. It is notable that many of the ideas presented here have been proposed in the past few years, and that dedicated R\&D, such as work that is planned in the SQMS center at Fermilab, has potential to make significant advances. In this context, these R\&D directions are synergistic with the quest to build SRF cavity-based quantum computers.

Many of the opportunities presented here are small-scale experiments.
In the long term, the most successful schemes may have room for growth and multiplexing. Modern SRF-based accelerators, such as PIP-II, LCLS-II, and the European XFEL are large integrated sytems with a large total cavity volume in excess of 10 m$^3$, and a total cooling power of order several kW. The well-motivated physics goals presented here may be significantly advanced by an efforts on this scale. A promising path towards ambitious science goals is to test search schemes with single-cavity setups in existing test stands. The most successful of these can then be scaled up.

\section{Acknowledgements}

This material is based upon work supported by the U.S. Department of Energy, Office of Science, National Quantum Information Science Research Centers, Superconducting Quantum Materials and Systems Center (SQMS) under the contract No. DE-AC02-07CH11359. Fermilab is operated by the Fermi Research Alliance, LLC under contract No. DE-AC02-07CH11359 with the United States Department of Energy.

DB is supported by a `Ayuda Beatriz Galindo Senior' from the Spanish `Ministerio de Universidades', grant BG20/00228.
IFAE is partially funded by the CERCA program of the Generalitat de Catalunya. The research leading to these results has received funding from the Spanish Ministry of Science and Innovation (PID2020-115845GB-I00/AEI/10.13039/501100011033). 

\bibliography{bibliography.bib}

\end{document}